\documentclass{article}
\usepackage{framed} 
\usepackage{nomencl}

\usepackage{classeJBArxive}

\newcommand{\degC}{^{\,\circ}C}

\usepackage{etoolbox}
\renewcommand\nomgroup[1]{%
  \item[\bfseries
  \ifstrequal{#1}{A}{Physics Constants}{%
  \ifstrequal{#1}{B}{Dimensionless symbols}{%
  \ifstrequal{#1}{C}{Boundary value problem symbols}{%
  \ifstrequal{#1}{D}{Approximation basis symbols}{%
  \ifstrequal{#1}{E}{PGD symbols}{%
  \ifstrequal{#1}{F}{Indicator}{%
  \ifstrequal{#1}{G}{Other Grec Symbols}{%
  \ifstrequal{#1}{H}{Other Latin Symbols}{}}}}}}}}%
]}

\title{Parametric PGD model used with orthogonal polynomials to assess efficiently the building's envelope thermal performance}
\author{.}
\author{Marie-Hélène Azam$^{1,2,*}$, Julien Berger$^{3}$, Sihem Guernouti$^{1,2}$, Philippe Poullain$^{1}$, Marjorie Musy$^{1,2}$}

\begin{document}

\maketitle
\begin{itshape}
$^1$ Université de Nantes, GeM UMR 6183 CNRS/Université de Nantes/Centrale Nantes, F-44600 Saint Nazaire, France\\
$^2$Cerema, Equipe de Recherche BPE, F-44000 Nantes, France\\
$^3$ Université de La Rochelle, LaSIE UMR 7356 CNRS, F-17000, La Rochelle, France \\
$^*$Corresponding author: Azam Marie-Hélène, marie-helene.azam@univ-nantes.fr, IUT de Saint-Nazaire, Département Génie Civil, 58 rue Michel Ange, F-44600 Saint Nazaire
\end{itshape}{}

\bigbreak
\textbf{Abstract:}
Estimating the temperature field of a building envelope could be a time-consuming task. The use of a reduced-order method is then proposed: the Proper Generalized Decomposition method. The solution of the transient heat equation is then re-written as a function of its parameters: the boundary conditions, the initial condition, etc.  To avoid a tremendous number of parameters, the initial condition is parameterized. This is usually done by using the Proper Orthogonal Decomposition method to provide an optimal basis. Building this basis requires data and a learning strategy. As an alternative, the use of orthogonal polynomials (Chebyshev, Legendre) is here proposed.     
\bigbreak
\bigbreak


\section{Introduction}
Modeling the thermal behavior of a building or a group of buildings is a challenging task. It implies that several physical phenomena have to be taken into account: short and long-wave radiative heat balance, sensible and latent heat flow transported by outdoor air movement and conductive heat transfer through the materials. All those heat fluxes may vary over time and through space and thus lead to complex and non-uniform boundary conditions. 
\bigbreak
To quantify the global heat loss of a building envelope, the balance between the outdoor thermo-radiative heat fluxes and the indoor ones must be estimated. One way of solving the global problem is to split the problem into several sub-problems relative to (i) the outside thermo-radiative balance, (ii) the inside thermo-radiative balance, and (iii) the heat transfer through the envelope \cite{malys2015microclimate}. Each problem is solved using a numerical model for each set of governing equations. The whole energy model represents then the aggregation of those several sub-models through a coupling procedure also called co-simulation \cite{berger2018intelligent}. Each numerical model exchanges parameters (\emph{i.e.} surface energy balance or surface temperature in the case of a thermal problem) with the other models during the simulation process.
\bigbreak
This combination of models results in a large computation complexity and we need to reduce the computational times. 

We focus here on the problem of the building envelope. To solve the global problem, for each element of the building (walls, floor, \emph{etc.}), the temperature field needs to be computed. For that purpose, the transient heat transfer equation needs to be solved for the previously described boundary conditions (indoor and outdoor thermo-radiative balance). Usually, a classical numerical model is then used based on finite difference, finite element, or finite volume. Those methods provide an accurate solution but for a high computation cost.
\bigbreak
To reduce the computational time keeping an accurate solution, the use of model order reduction methods is currently investigated. The main idea is to replace the detailed and time-consuming model with a reduced-order model. For that purpose, we investigate the use of the Proper Generalized Decomposition (PGD) method.
\bigbreak
Applied to urban soil heat transfer modeling, this model reduction method has shown its efficiency \cite{azam2018mixed}. A cut computational cost of $80\%$ was observed for a mean surface temperature error below  $0.52 \degC$. Applied to building wall heat transfer modeling, the PGD parametric model computes the solution 100 times faster than a classical numerical method \cite{berger2017innovative}. 
\bigbreak
To reduce the numerical complexity of the problem, the solution is decomposed as a function of parameters like the boundary conditions, or the initial condition. The efficiency of the PGD method relies on the number of parameters used. To obtain a minimum number of parameters, some of them are usually combined through approximation. Those approximations are done by the projection of the field of interest on an approximation basis. 
\bigbreak
Selecting the right approximation basis that will guarantee the model final accuracy with a minimum number of parameters is a challenging task. The purpose of this article is then to overcome this obstacle.
We investigate here the use of a polynomial basis like \textsc{Chebyshev} or \textsc{Legendre}. Out of the approximation theory \cite{trefethen2013approximation}, those basis have proven to be very efficient at solving partial differential equations using spectral methods \cite{gasparin2018solving,gasparinspectral}.
\bigbreak
The use of a polynomial basis is compared to the use of a classical reduced-order basis obtain through the Proper Orthogonal Decomposition (POD) method. For that purpose, section \ref{sec:methods} presents each basis and their combination with the PGD method. Each combination is compared both on its accuracy and computation time. The global methodology applied is explained in section \ref{sec:methodology}. To evaluate the basis in several situations, two case studies are presented. The first case study, presented in section \ref{sec:theoretical_case_study} is a theoretical application. In section \ref{sec:Practical application}, the models are then applied to a practical case with realistic boundary conditions. The results of the models will be confronted with laboratory measurements. For each case study, the influence of several parameters on the accuracy of the approximation basis is investigated: the number of modes in the approximation basis, the number of modes in the PGD model and the discretization.
\begin{table*}[!t]
       \begin{framed}
      \footnotesize
         \printnomenclature
       \end{framed}
\end{table*}

\section{Materials and Methods}
\label{sec:methods}

\subsection{Physical problem of heat transfer in building wall}
\label{sec_2_1:Physical_problem}

The physical problem studied is defined to be as close as possible to problems usually solved by building energy models (e.g. EnergyPlus \citep{crawley2001energyplus}). It involves transient one-dimensional heat conduction through a wall without volumetric heat dissipation for a time interval $\Omega_{\,\tau}$ with $ \mathrm{t \in \, \bigl[\, 0 \,,\, \tau \, \bigr] }$ and space interval $\Omega_{\,\mathrm{x}}$with $ \mathrm{x \in \, \bigl[\, 0 \,,\, L \, \bigr] }$:
\nomenclature[A]{$\mathrm{L}$}{Length of the wall [$\,\mathsf{m}\,$]}
\nomenclature[A]{$\tau$}{Simulation time horizon}
\nomenclature[A]{$\Omega$}{Domain definition}
\begin{equation}
\mathrm{c \, \frac{\partial u (x,t)}{\partial t}  \ = \  \frac{\partial}{\partial  x} \, \biggl(\, k \, \frac{\partial u (x,t)}{\partial x}} \, \biggr) \,,
\label{eq:HTE}
\end{equation}
\nomenclature[A]{$\mathrm{u}$}{Temperature field[$\,\mathsf{K}\,$]}
\nomenclature[A]{$\mathrm{k}$}{Thermal conductivity [$\,\mathsf{W}.\mathsf{m}^{-1}.\mathsf{K}^{-1}\,$]}
\nomenclature[A]{$\mathrm{c}$}{Volumetric heat capacity [$\,\mathsf{J}.\mathsf{m}^{-3}.\mathsf{K}^{-1}\,$]}

On each side of the wall, a \textsc{Fourier} boundary condition is assumed.
On $\mathrm{x}  \ = \  0$, the boundary condition can be described by the following equation:  
\begin{align}
\mathrm{- \, k \ \frac{\partial u (x,t)}{\partial x}   \ = \  q (t)  \ - \   h_{\, out} \, \Bigl(\, u (x,t) \ - \ u_{\,out} (t)} \, \Bigr) \,, && \mathrm{x}  \ = \  0 \,,
\end{align}
\nomenclature[A]{$\mathrm{h}$}{Convective heat transfer coefficient [$\,\mathsf{W}.\mathsf{m}^{-2}.\mathsf{K}^{-1}\,$]}
\nomenclature[A]{$\mathrm{q}$}{Net radiative heat flux [$\,\mathsf{W}.\mathsf{m}^{-2}\,$]}
\nomenclature[A]{$\mathrm{u_{\,out}, \, u_{\,in}}$}{Outside and inside air temperature [$\,\mathsf{K}\,$]}
The surface energy balance depends on a net radiative heat flux, noted $\mathrm{q}$, and a sensible heat flux. The last one is calculated from the outdoor air temperature $\mathrm{u_{out}}$ varying over time and from a convective heat transfer coefficient $\mathrm{h_{\, out}}$. 
\bigbreak
On $\mathrm{x  \ = \  L}$, the boundary condition can be described by the following equation:
\begin{align}
\mathrm{ k \ \frac{\partial u(x,t)}{\partial x}   \ = \  \ - \ h_{\, in} \, \Bigl(\, u(x,t)  \ - \  u_{\, in }(t) \, \Bigr) \,,} && \mathrm{x  \ = \  L} \,.
\end{align}
As it is illustrated with the outside boundary condition, another radiative heat flux could have been added to the inside boundary condition to complexify the mathematical model proposed here. The net radiative heat flux is neglected on that side of the wall. To support this hypothesis, the error due to this simplification of the mathematical model is studied in the appendix \ref{app:details_on_the_model_error}. Note that this assumption will not have an impact on the results presented because the same mathematical model is used for all the numerical models developed.

The sensible heat flux is calculated from the indoor air temperature $\mathrm{u_{in}}$ that varies over time and from a convective heat transfer coefficient $\mathrm{h_{\, in}}$.

The initial temperature is uniform:
\begin{align}
\mathrm{u(x,t)  \ = \  u_{\,0}  \,, } && \mathrm{t  \ = \ 0} \,.
\end{align}
\bigbreak 
Equation (\ref{eq:HTE}) can be written in a dimensionless form as: 
\begin{equation}
 \frac{\partial u(x,t)}{\partial t} \ = \ Fo \, \frac{\partial^{\,2}  u(x,t)}{\partial x^{\, 2}} \,,
 \label{Eq:EqChaleurAdim}
\end{equation}for a time interval $\Omega_{\,\Gamma} \,  \ = \  \, \bigl[\, 0 \,,\, \Gamma \, \bigr]$ and space interval $\Omega_{\,x} \,  \ = \  \, \bigl[\, 0 \,,\, 1 \, \bigr]$,
\nomenclature[B]{$\Gamma$}{Dimensionless simulation time horizon}
and the boundary condition as:
\begin{subequations}
\begin{align}
\ \frac{\partial u (x,t) }{\partial x}   & = \  Bi_{out} \, \Bigl(\, u \ - \ u_{\, out} \, \Bigr) \ - \ q \,, && x  \ = \  0 \,, \label{Eq:BC_ext_Adim} \\[4pt]
\ \frac{\partial u (x,t) }{\partial x}   & = \ \ - \  Bi_{in} \, \Bigl(\, u \ - \  u_{\, in} \, \Bigr)\,, && x \ = \ 1 \,. \label{Eq:BC_in_Adim}
\end{align}
\end{subequations}
The initial condition becomes:
\begin{align}
u \ = \  0,  && t \ = \ 0\, .
\label{Eq:IC_Adim}
\end{align}
Where the dimensionless quantities are defined as: 
\begin{equation*}
u : \ = \ \mathrm{\frac{u \ - \ u_{\,0}}{u_{\,0}} }; \; \; \;
\, t : \ = \ \mathrm{ \frac{t}{t_{\, ref}} }; \; \; \;
\, x  \ = \ \mathrm{ \frac{x}{L} } ; \; \; \;
\, Bi_{\,in} : \ = \ \mathrm{ \frac{h_{in}.L}{ k} } ; \; \; \;
\, Bi_{\,out} : \ = \ \mathrm{ \frac{h_{out}.L}{ k} }; \; \; \;
\, Fo : \ = \ \mathrm{ \frac{ k \ .  \, t_{ref} }{c \, L^2} } \ = \ 1
\end{equation*}
\begin{equation*}
\, \mathrm{ t_{\,ref} } : \ = \ \mathrm{ \frac{c \, L^2}{ k  } } ; \; \; \;
\, u_{\, in} : \ = \ \mathrm{ \ - \ 1 \ + \  \frac{u_{\,in}}{u_{0}} }; \; \; \;
\, u_{\, out} : \ = \ \mathrm{ \ - \ 1 \ + \  \frac{u_{\,out}}{u_{0}} }; \; \; \;
\, q  : \ = \ \mathrm{ \frac{q \, . \, L }{   k \,  . \, u_{0} } } ; \; \; \;
\, \Gamma = \mathrm{ \frac{\tau}{t_{\, ref}} }
\end{equation*}
\nomenclature[B]{$Bi$}{Biot Number}
\nomenclature[B]{$Fo$}{Fourier Number}
\subsection{The related boundary value problem in the context of co-simulation}

\begin{figure}[htp]
    \centering
    \includegraphics[width=0.46\textwidth]{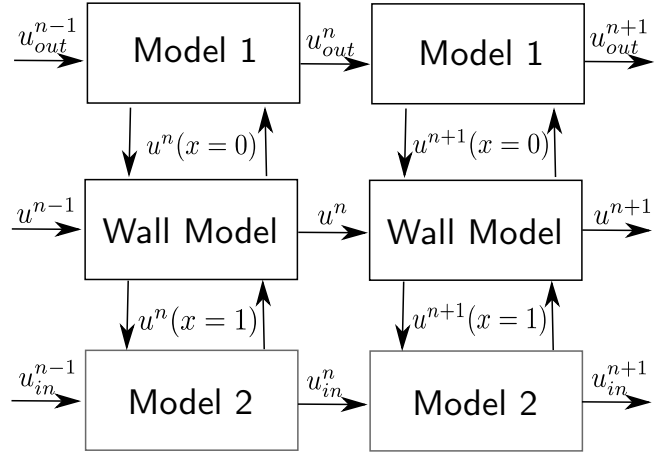}
    \caption{Co-simulation process}
    \label{fig:co-simulation}
\end{figure}

The physical problem involves the partial differential equation (PDE) Eq.~\eqref{Eq:EqChaleurAdim} together with the boundary (Eqs.~\eqref{Eq:BC_ext_Adim} and \eqref{Eq:BC_in_Adim}) and initial conditions (Eq. \eqref{Eq:IC_Adim}). As presented on Figure \ref{fig:co-simulation}, it is solved in the context of co-simulation (or coupling) with other numerical models (models 1 and 2), by solving the radiative heat balance and the air transfer around the building walls.

In this context, the initial boundary value problem Eq.~\eqref{Eq:EqChaleurAdim} is semi-discretized along the time line \cite{gasparin2018adaptive}. The time discretization parameter is denoted by $\Delta t\,$, corresponding to the time step of coupling between the numerical models of the co-simulation. The discrete values of functions $u \, (\,x\,,\,t\,)$ is written as $u^{\,n} \ \eqdef \ u\,(\,x\,,\,t^{\,n}\,)$ with $n \ = \ 1 \,, \ldots \,, N_{\,t} \,$. Thus, using an implicit approach, Eq.~\eqref{Eq:EqChaleurAdim} becomes:
\begin{equation}
\label{eq:chaleur_semidiscret}
u^{\,n + 1}  \ = \ u^{\,n} \ + \Delta t \cdot Fo \cdot  \frac{\partial^{\,2} u^{\,n + 1}}{\partial x^{\,2}} \,,
\end{equation}
By introducing $y \, \equiv \, u^{\,n+1}\,$, Eq.~\eqref{eq:chaleur_semidiscret} can be reformulated as:
\begin{equation}
\label{eq:BVP}
y \ - \ a \cdot \frac{\partial^{\,2} y}{\partial x^{\,2}} \ = \ b\,(\,x\,) \,.
\end{equation}
Here, $y$ is the unknown of our boundary value problem and depends on the space coordinate $x\,$. The coefficient $a \, \eqdef \, \Delta t \cdot Fo$ depends on the properties of the material composing the wall and on the co-simulation time step. The coefficient $b \, \eqdef \, u^{\,n}$ is qualified as the source term of the boundary value problem, depending on the space coordinate $x\,$. It also varies at each time step of the co-simulation. The boundary conditions Eqs.~\eqref{Eq:BC_ext_Adim} and \eqref{Eq:BC_in_Adim} are also transformed:
\begin{subequations}
\label{eq:BVP_BC}
\begin{align}
\ \frac{\partial y}{\partial x}   & = \,  Bi_{out} \cdot y \ - \ b_{\, out} \,, && x  \ = \  0 \,, \\[4pt]
\ \frac{\partial y}{\partial x}   & = \, - \,  Bi_{in} \cdot y \ + \  b_{\, in} \,, && x \ = \ 1 \,, 
\end{align}
\end{subequations}
where the coefficients $b_{\, out}$ and $b_{\, in}$ are:
\begin{align*}
b_{\, out} & = \, - \, Bi_{out} \cdot u_{\, out} \,(\,t^{\,n}\,) \ - \ q\,(\,t^{\,n}\,)   \,, 
&& b_{\, in} \ = \ Bi_{in} \cdot u_{\, in} \,(\,t^{\,n}\,) \,.
\end{align*}
Both are constants given at each time step $\Delta t$ of the co-simulation by model 1 and 2.

\subsection{Formulation of the parametric problem}
\label{sec:Parametric_formulation}
The boundary value problem Eq.~\eqref{eq:BVP} together with the boundary conditions \eqref{eq:BVP_BC} are the main interest to build a reduced-order model. Several solvers exist to solve such a problem. A brief overview can be consulted in \cite{shampine2000solving}. These numerical models are used to compute a solution  $y\,(x\,)\,$ only depending on the space coordinate. 
\bigbreak
However, it is a challenging problem to build a solution depending on the space coordinate and on extra-parameters such as the source term $b$ and the coefficients $b_{\, out}$ and $b_{\, in}\,$. It requires to solve a so-called parametric problem. The use of the PGD methods gives the opportunity to decompose the solution of a problem as a function of any parameters to generate a parametric model. 
\bigbreak
Taking into account the source term as a parameter is another challenging task. Indeed, once discretized in space, the source term is made of discrete values: one information per piece of the mesh. It implies inputting as many parameters in the parametric model as the number of pieces of the mesh. To avoid this large number of involved parameters, the source term is approximated by its projection on an approximation basis with a lower rank:
\begin{equation}
	b \, \bigl(\,x\,\bigr) \ = \ \sum_{j \ = \ 1}^{\,\mathcal{N}} \Psi_{\,j} \, \bigl(\,x\,\bigr) \, \zeta_{\, j}
\end{equation}
where $\Psi_{\,j}$ is the approximation basis, $\zeta_{\,j}$ are the coefficients of the projection and $\mathcal{N}$ the number of modes in the basis. So, the solution of Eq.~\eqref{eq:BVP} is searched as:
\begin{align*}
u \,:\,   \Bigl[\, 0 \,,\, 1 \, \Bigr] \times \Omega_{\,b_{\,out}} \times \Omega_{\,b_{\,in}} \times \Omega_{\,\zeta_{\,j}} & \longrightarrow \, \mathbb{R}  \,,  \\[4pt]
\bigl(\,x\,,\,b_{\, out}\,,\,b_{\, in}\,,\, \zeta_{\,j}\,\bigr) & \longmapsto \, u\;\bigl(\,x\,,\,b_{\, out}\,,\,b_{\, in}\,,\, \zeta_{\,j}\,\bigr) \,.
\end{align*}
The sets $\Omega_{\,b_{\,out}}\,$, $\Omega_{\,b_{\,in}}$ and $\Omega_{\,\zeta_{\,j}}$ are the domain of variations of the coordinates $b_{\, out}\,$, $b_{\, in}$ and $\zeta_{\,j}\,$, respectively. They are defined such as: 
\begin{align*}
\Omega_{\,b_{\,out}} & \ = \  \bigl[\,b_{\,out}^{\,-} \,,\, b_{\,out}^{\,+} \,\bigr] \,, 
&& \Omega_{\, b_{\, in}} \ = \ \bigl[\,b_{\,in}^{\,-} \,,\, b_{\,in}^{\,+} \,\bigr]\,,
&& \Omega_{\, \zeta_{\,j}} \ = \ \bigl[\,\zeta_{\,j}^{\,-} \,,\, \zeta_{\,j}^{\,+} \,\bigr] \,,
&& j \in \bigl[\,0 \,,\, \mathcal{N} \,\bigr] \,.
\end{align*}
Their respective discretization parameters are denoted by $\Delta_{\,b_{\,out}}\,$, $\Delta_{\,b_{\,in}}$ and $\Delta_{\,\zeta_{\,j}}\,$.

\nomenclature[C]{$n$}{Temporal subscript}
\nomenclature[E]{$\mathcal{M}$}{PGD model number of modes}
\nomenclature[D]{$\Psi$}{Approximation basis}
\nomenclature[D]{$\zeta$}{Approximation coefficients}
\nomenclature[D]{$\mathcal{N}$}{Approximation basis number of modes}

\nomenclature[C]{$b$}{Source term}
\nomenclature[C]{$a$}{Constant in the ODE}
\subsection{Approximation basis}
\label{sec_2_4:Approximation_basis}
In the literature, several parameterizations have been studied. Chinesta \emph{et al.} (2013 \cite{chinesta2013proper}) and Gonzalez \emph{et al.} (2012 \cite{gonzalez2012proper}) proposed to use the nodal values corresponding to the piece-wise linear finite element approximation of the problem. However, according to Gonzalez \emph{et al.} (2014 \cite{gonzalez_real-time_2014}), this method leads to a large number of degrees of freedom: their model is made of one parameter per nodal values.  That is why they proposed to use the POD to provide a suitable parameterization of the initial condition with the lowest number of degrees of freedom \cite{gonzalez_real-time_2014}. More information can be found on this method applied to convective heat transfer in \cite{zucatti2020assessment} and solid dynamics in \cite{cueto_proper_2016,gonzalez_real-time_2014}. 
\bigbreak
One of the main drawbacks of this method is that a learning process is needed. It has an impact on the accuracy of the reduced-order basis. For this reason, the data-set used must be representative of the problem (boundary values, initial conditions, materials used). 
\bigbreak
According to Gonzalez \emph{et al.} (2014 \cite{gonzalez_real-time_2014}) the initial condition could be interpolated by piece-wise polynomials. However, for the specific field of solid dynamics, this approach is not the best choice, considering the behavior of the system. Another solution proposed by Poulhaon \emph{et al.} (2012, \cite{poulhaon2012first}) is to use an auxiliary mesh much coarser than the one used for the solution of the problem. A projection is made from the fine to the coarse mesh using the least square method. This method is purely mathematical and does not take into account physical considerations such as energy conservation or heat flux conservation. According to Poulhaon \emph{et al.} (2012, \cite{poulhaon2012first}) it should be completed by a mathematical tool to take into account the physics of the studied phenomenon.
\bigbreak
Conforming to \cite{borzacchiello2017non}, one important feature for the choice of an approximation basis is the \textit{sparsity}. It ensures that the chosen basis has the required regularity to represent the solution. Spectral basis, such as polynomial or trigonometric functions, guarantee sparsity. For such functions, the values of the coefficients decrease exponentially with the order of approximation \cite{boyd2001chebyshev}. However, the basis is full (because a spectral basis is not interpolative \cite{borzacchiello2017non}). It implies that the computational cost needed to determine the coefficients becomes impractical for large systems.
\bigbreak
Based on the literature review, two methods are here compared: the use of a polynomials basis and the use of a POD basis to approximate the temperature profile. Details on how to build each approximation basis used are given in the appendix \ref{app:details_on_the_app_basis_cons}.
\bigbreak
As stated by the \textsc{Weierstrass} approximation theorem, every continuous function on a bounded interval can be approximated by a polynomial to a certain accuracy \cite{trefethen2013approximation}. Several functions with polynomial basis can then be used according to the studied problem. The most simple polynomial basis is the monomial one. As described by Peyret (2013 \cite{peyret2013spectral}), if a periodic problem is studied, the \textsc{Fourier} method should be used. Yet, this method is not suitable for non-periodic problems, because of the \textsc{Gibbs} phenomenon. In this case, orthogonal polynomials such as \textsc{Chebyshev} or \textsc{Legendre} polynomials should be used.
\bigbreak 
Considering the numerous polynomial basis, the first difficulty is to select the right basis for the considered problem. As the field of interest is a non-periodic, smooth function, \textsc{Fourier} or \textsc{Laurent} polynomials shall not be studied here.
\bigbreak
According to Trefethen (2013, \cite{trefethen2013approximation}), the monomial basis is comfortable but should never be used to approximate a function. If we compare the condition number for inversion of the three basis, the \textsc{Chebyshev} and \textsc{Legendre} polynomials basis have a smaller condition number than the monomial one. If the condition number of a matrix is large, the matrix is close to being singular. The condition number reveals that the projection of the field of interest on the monomial basis will be sensitive to numerical round-off errors and perturbations in the input data. Moreover, monomial basis do not meet \textit{sparsity} condition as its coefficients increase with the order. Therefore, this basis should not be used here to parameterize the initial condition. 
\bigbreak 
According to Trefethen (2013, \cite{trefethen2013approximation}), \textsc{Legendre} points and polynomials are neither better than \textsc{Chebyshev} ones for approximating functions, nor worse. The main advantage to use \textsc{Chebyshev} over \textsc{Legendre} points center around the use of FFT (Fast \textsc{Fourier} Transform). This function can be used to get the coefficients from the point values or the reverse. But this property is not used here. Both polynomials basis will be compared.
\bigbreak
The \textsc{Chebyshev} and \textsc{Legendre} polynomials are part of the family of orthogonal polynomials. They are calculated respectively at the \textsc{Chebyshev} and \textsc{Legendre} points. Special attention must be given to the spatial domain of the problem. The points define a non-uniform mesh for a space interval $[-1,1]$. Thus, a change of variable must be performed to transform the dimensionless spatial domain $[0,1]$ to $x \in [-1,1]$. 

\subsection{Proper Generalized Decomposition method}
\label{sec:PGD}

Several MOR methods can be used to solve a parametric problem. One of them is the Proper Generalized Decomposition Method (PGD). It is an \textit{a priori} MOR method based on the separation of variables. It does not reduce the system of equations itself but the whole parametric problem. Any variable can then be defined as an extra-parameter of the model \cite{borzacchiello2017non}.
\bigbreak
With spectral methods \cite{gasparin2018advanced}, the PGD method is one of the unique methods that allows to create a complete parametric model without knowing \textit{a priori} the solution of the problem.
\bigbreak
The PGD is used to propose an accurate parametric solution of the formulated BVP problem. The method approximates the solution as a finite sum of separable functions. As presented in Section \ref{sec:Parametric_formulation}, the parametric model involves three parameters: the space, the boundary condition and the source term. Applying the PGD method, the solution is sought as the sum of $\mathcal{M}$ functional products involving each function as follows:
\begin{equation}
y  \ = \  \sum_{m \ = \  1}^\mathcal{M} X_{\,m} \,\bigl(\,x\,\bigr) \, E_{\,m} \,\Bigl(\,b_{in}\,\Bigr) \, F_{\,m} \,\Bigl(\,b_{out}\,\Bigr)  \, \prod_{j  \ = \  1}^\mathcal{N} G_{\,m}^{j}\,\bigl(\, \zeta^{j}\,\bigr)
\label{eq:SolPGD_param}
\end{equation}
where $X$, $E$, $F$, and $G$ designate the functions of the parameters. Each function is defined over a domain : $\Omega_x \, = \, \bigl[\, -1, 1 \, \bigr]$, $ \Omega_{b_{in}} \, = \,\bigl[\,b_{in}^{-}, b_{in}^{+}\, \bigr] $, $ \Omega_{b_{out}} \, = \,\bigl[\,b_{out}^{-}, b_{out}^{+}\, \bigr] $ and $\Omega_{\zeta_j} \, = \, \bigl[\zeta_{\,j}^{-}, \zeta_{\,j}^{+} ]$. 
\bigbreak
The following weak form of the ODE is used with the test function $y^*$ (Galerkin formulation):
\begin{equation} 
\int_{\Omega_x \times \Omega_{b_{in}} \times \Omega_{b_{out}} \times \Omega_{\zeta_j} } y^*.  \left(
    y \ - \ a \, \frac{\partial^{\, 2} y }{\partial x^{\, 2}} \ - \ \sum_{j \ = \ 1}^{\,\mathcal{N}} \Psi_{\,j} \, \bigl(\,x\,\bigr) \, \zeta_{\, j}\right)  \, \mathrm dx . \, \mathrm d b_{in}. \, \mathrm d b_{out} .\, \mathrm d \zeta_j  \ = \ 0
\end{equation}
The weak form of the ODE is regarded as an optimization problem. It leads to a nonlinear optimization problem due to the functional product of the subspaces. It can be solved with an iterative procedure that features two nested loops: the alternating direction strategy and the enrichment process \cite{chinesta2011short}. The calculation of the unknowns is performed alternatively along each dimension until convergence \cite{ammar2010error}. In this way, the algorithm splits the high dimensional problem into a series of low dimensional ones. The complexity of the problem then grows linearly with the number of parameters. 
Each function $X_{\,m}$, $E_{\,m}$, $F_{\,m}$ and $G_{\,m}^j$ is first randomly initialized and then solved by iterations. The alternating directions process stops once a fixed point is reached. The criterion $\tilde{\epsilon}$ used to make this determination is defined by the user \cite{chinesta2013proper}. Once this criterion is reached, the new functions are added to the previous one in the PGD basis. The enrichment process of the PGD basis stops when the $\epsilon$ criterion, defined by the user, is reached \cite{chinesta2013proper}. Details on the alternating directions strategy equations and algorithms for a similar problem can be found on \cite{azam2018mixed}. For further details on the method and its developments, the interested reader may refer to \cite{chinesta2013proper,cueto_proper_2016}. 
\nomenclature[E]{$\tilde{\epsilon}$}{Fixed point criterion}
\nomenclature[E]{$\epsilon$}{Enrichment process stopping criterion}
\nomenclature[E]{$\Delta$}{Discretization of the parameter}
\bigbreak
Each function ($X_m$, $E_m$, $F_m$, $G_m^{j}$) defined previously depends on a continuous variable. To solve the parametric problem with the previous algorithm, the continuous variables need to be discretized. For that purpose, the continuous variable is projected on a mesh. The continuous variable is then described by a vector. The finer the mesh of discretization of each parameter, the closer the discrete value to the continuous one. But as a results, the number of elements in the vectors used to describe the parameter increases. 
\bigbreak 
According to Leon \emph{et al.} (2018, \cite{leon2018wavelet}), the final accuracy of a PGD model depends on the number of terms $\mathcal{M}$ in the final sum, on the number of parameters/vectors ( $x$, $b_{in}$, $b_{out}$, $\zeta_j$) and the discretization of those parameters. However, by increasing the number of elements in the mesh of discretization for each parameter, we increase the complexity of the problem. In the case of a PGD model, this complexity grows linearly with the number of parameters \cite{ammar2010error,borzacchiello2017non}. As a comparison, the complexity of a grid-based discretization (finite element, finite difference) grows exponentially with the number of mesh elements. The number of elements on each vector is a matter of CPU time and space to save the PGD parametric model. As the purpose of building a PGD parametric model is to decrease the calculation time (compared to a classical model: finite difference, finite element) the number of elements in each vector should be then optimized.
\bigbreak 
For the spatial parameter and the boundary condition, the methodology to defined the discretization is classical, no special interrogation arises. However, each mode of the approximation basis also needs to be discretized. Several questions can arise for the coefficients $\zeta_j$ of the source term approximation. Spectral basis such as \textsc{Chebyshev} or \textsc{Legendre} guarantee sparsity. When this condition is met, the order of magnitude of the coefficients $\zeta_j$ decreases exponentially with the order of approximation \cite{boyd2001chebyshev}. The discretization of each coefficient needs to fit the order of magnitude of each mode. To simplify our study and only use one parameter to define the discretization of each coefficient of the basis, we propose to use dimensionless numbers for the coefficients $\zeta_j$ defined as $\overline{\zeta_j}$.

\begin{equation}
    \overline{\zeta_j} = \frac{\zeta_j - min \left( \zeta_j \right) }{ max \left( \zeta_j \right) - min \left( \zeta_j \right)}
\end{equation}

where $\overline{\zeta_j} \in [0, 1]$ and $\zeta_j \in [ min \left( \zeta_j \right), max \left( \zeta_j \right)]$. 

\subsection{\textit{Offline}/\textit{online} strategy}
\label{Sec:offline_online}
\begin{figure}[htp]
    \centering
    \includegraphics[width=1.0\textwidth]{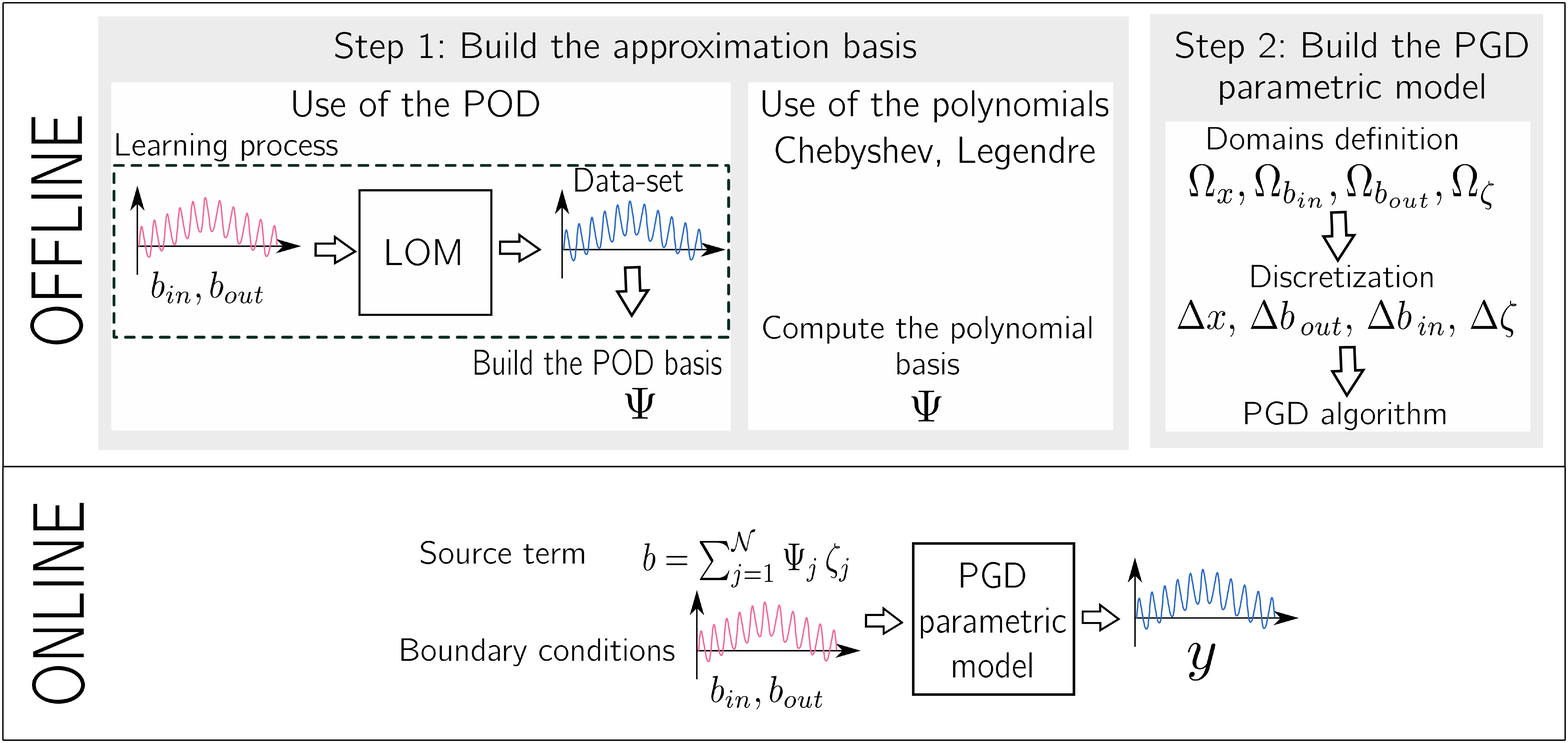}
    \caption{\textit{Offline}/\textit{online} strategy}
    \label{fig:offline_online}
\end{figure}

The use of the PGD method, to solve the parametric problem, features an offline-online strategy. During the offline stage, the model is built for the set of parameters. It is then used online combined with other models. Online, the use of the model requires no more than reading the unknown value in an abacus. 
\bigbreak
As previously described, one of the parameters of the problem consists of the source term $b$. Taking into account the source term as a parameter is a challenging task. Indeed, once discretized in space, the source term is made of discrete values: one information per piece of the mesh. It implies inputting as many parameters in the PGD parametric model as the number of pieces of the mesh, plus the boundary conditions and spatial coordinates. The PGD method has shown success for problems up to dimension 100. However, the efficiency of a parametric model depends on the number of involved parameters  \cite{gonzalez_real-time_2014, poulhaon2012first}. 
\bigbreak
To avoid this large number of involved parameters, one can gather some of them. Considering the source term, this is usually done by using an approximation basis. The temperature field is projected on an approximation basis of a smaller size. The use of different approximation basis is investigated in this work: \textsc{Chebyshev}, \textsc{Legendre} polynomials and the POD reduced basis. 
\bigbreak
The PGD method is combined with the approximation basis to build a PGD parametric model for the previous presented physical problem. Each step of the \textit{offline}/\textit{online} strategy is described in Figure \ref{fig:offline_online}.
\bigbreak
The first step of the \textit{offline} phase consists of building an approximation basis. In the case of \textsc{Chebyshev}, \textsc{Legendre} basis, it is made of the polynomials. In the case of the POD reduced basis, a learning process is required. The POD basis is built on a data-set. The latter can be provided from available measurements or from another model defined as a Large Original Model (LOM). To get an accurate approximation basis, the learning process needs to be representative of all future modeled combinations. In the specific case of a building energy model, the basis should be representative of every material and climate data that could be used. The learning process needs a large amount of data and could be very time-consuming.
\bigbreak 
The approximation basis $\Psi$ aims at representing the source term in a minimum number of parameters called modes. For that purpose, the approximation basis is truncated. A number of modes in the approximation basis, $\mathcal{N}$, is defined to achieve the desired approximation accuracy. Note that this number has a direct influence on the number of parameters used in the PGD parametric model and its accuracy.
\bigbreak 
Then, all the parameters of the model (the mesh, the boundary conditions, and the approximation basis modes) are converted into parameter vectors. The discretization ($\Delta x, \, \Delta b_{\, out}, \, \Delta b_{\, in}, \, \Delta \zeta$) selected for each vector has an impact on the accuracy of the PGD parametric model. 
\bigbreak 
Finally, as a last step of the \textit{offline} phase, the parametric problem can be solved with the PGD algorithm. The PGD parametric model is built for a number of PGD modes $\mathcal{M}$. This parameter also influences the accuracy of the parametric model. 
\bigbreak 
Once the PGD parametric model has been built, it can be applied for any value within the previously defined intervals, \textit{online}. The source term $b$ is projected on the approximation basis $\Psi $ to identify the parameters $\zeta^{j}$. Afterwards, the PGD modes are computed for the defined parameters $x$, $b_{in}$, $b_{out}$ and $\zeta^{j}$. The evaluation of the solution demands no more than reading a look-up table \cite{borzacchiello2017non}.

\section{Methodology}\label{sec:methodology}

The purpose of this article is to overcome the obstacle of parameterizing the initial condition of a PGD parametric model.
It is then necessary to quantify and compare the accuracy of each approximation basis in the framework of their combination with the PGD. The proposed study will therefore cover several issues:

\begin{enumerate}
    \item the accuracy of the approximation basis for a given number of modes $\mathcal{N}$,
    \item the discretization of each of the parameters vectors,
    \item the number of PGD modes $\mathcal{M}$.  
\end{enumerate}
    
For the use of the POD approximation basis, a supplementary issue has to be added: the efficiency of the learning process.

\subsection{Methods assessment's procedure}
\label{sec:Methods assessment's procedure}

To evaluate the approximation basis in several situations, two case studies are presented. The first case study is a theoretical application. It is used to study the influence of the three first issues cited previously. 
\bigbreak
The built basis are then applied to a practical case with realistic boundary conditions. The results of the models will be confronted with laboratory measurements. The influence of the learning period is studied through this second case study.
\bigbreak
They may seem simple and we could have considered more complicated case studies. However, the parametric model would have been more complex. It would then have been more complicated to identify the influence of the studied parameters on the final error of the model. 
\bigbreak
For each case study, the global methodology consists of two main steps. First, the approximation of the source term is evaluated to study the behavior of the basis alone. Then the PGD parametric model is evaluated to verify if the basis have the same behavior once applied in the PGD framework.  The performance of the three basis is compared with regards to the model errors and CPU time. The chosen indicators are presented hereafter.

\subsection{Error indicator of the model}
For each step of the assessment procedure, the error indicator chosen is the $ \ell_\infty$ norm. It is computed as the Root Mean Square Error between two spatial profiles. Only the maximum of the previous function is observed. This Section describes the errors calculated for each of the three parameters studied in this paper.

\subsubsection{Evaluation of the source term approximation}
\begin{figure}[htp]
    \centering
    \includegraphics[width=0.5\textwidth]{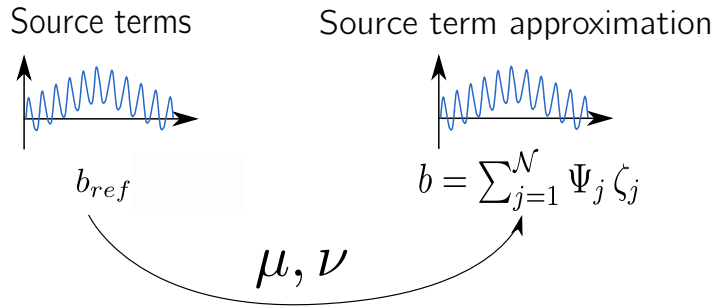}
    \caption{Evaluation of the source term approximation}
    \label{fig:methodology_metrics1}
\end{figure}
For each metrics introduced hereafter, Figure \ref{fig:methodology_metrics1} summarizes the methodology. First, the performance of each basis to approximate the source terms is evaluated by projecting the source terms (actual $b$ and reference $b_{ref}$) on the different basis and by then calculating the errors $\mu$ as follows: 
\begin{equation}
    \mu \, : ( \, \mathcal{N}, \, \Psi ) \mapsto\ \max _t \left(\, \sqrt{ \frac{1}{\mathtt{N_x}} \sum _{0}^{\mathtt{N_x}} \left [ b_{ref} -  \sum_{j \ = \ 1}^{\,\mathcal{N}} \Psi_{\,j}  \, \zeta_{\, j} \right ]^2  } \,\right)
\end{equation}
\nomenclature[F]{$\mu$}{Error due to the approximation by the basis}
where $\mathtt{N_x}$ is the number of elements over the axis. The reference source term (noted  $b_{ref}$ ) outcomes from the reference solution calculation at each point of the spatial mesh and for each studied time step.  The error is calculated for each approximation basis $\Psi$. The influence of the parameter $\mathcal{N}$ on the accuracy of the basis $\mu$ will be studied.
\bigbreak
To integrate the approximation basis into the PGD framework, each parameter of the model has to be discretized. The error due to this discretization noted $ \nu $ is evaluated for each approximation mode. The error is calculated as follows for the PGD variable $\zeta$: 
\begin{equation}
    \nu : (\, \mathcal{N}, \, \Psi, \, \zeta, \, \Delta \zeta ) \mapsto\ \max _t \left(\, \sqrt{ \frac{1}{\mathtt{N_x}} \sum _{0}^{\mathtt{N_x}} \left [b_{ref} -  \sum_{j \ = \ 1}^{\,\mathcal{N}} \Psi_{\,j}  \, \overline{\zeta_{\, j}} \right ]^2  } \,\right)
\end{equation}
with $\overline{\zeta}$ the dimensionless coefficients.
\nomenclature[F]{$\nu$}{Error due to the discretization of the parameters}

\subsubsection{Evaluation of the PGD parametric model}
\begin{figure}[htp]
    \centering
    \includegraphics[width=0.8\textwidth]{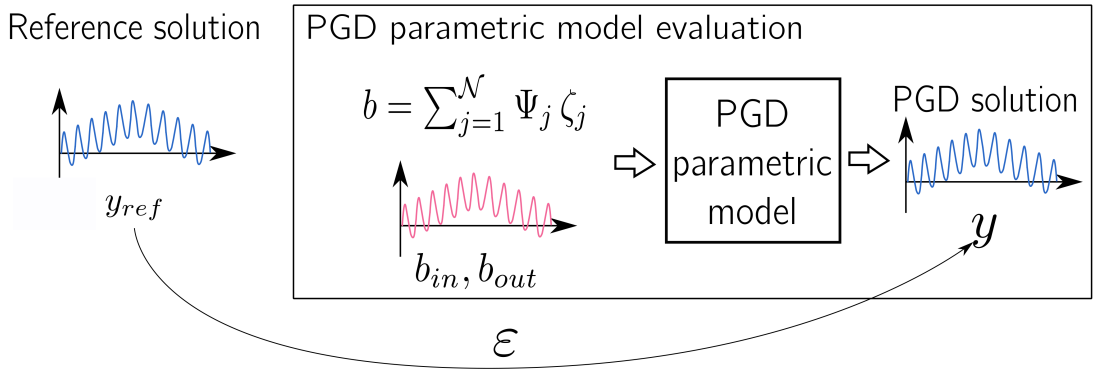}
    \caption{Methods assessment's procedure sum-up}
    \label{fig:methodology_metrics2}
\end{figure}
Finally, the approximation basis are introduced into a PGD parametric model to get a combined parametric model. The error of the combined model noted $\varepsilon$ is computed between the calculated temperature profile and the reference solution. Figure \ref{fig:methodology_metrics2} summarizes the methodology.
\begin{equation}
    \varepsilon : \, \mathcal{N}, \, \zeta, \, \Psi,\, \Delta \zeta , \, \mathcal{M} \mapsto\ \max _t \left(\, \sqrt{ \frac{1}{\mathtt{N_x}} \sum _{0}^{\mathtt{N_x}} \left [ y_{\, ref} -  \sum_{m \ = \  1}^\mathcal{M} X_{\,m} \,\bigl(\,x\,\bigr) \, E_{\,m} \,\Bigl(\,b_{in}\,\Bigr) \, F_{\,m} \,\Bigl(\,b_{out}\,\Bigr)  \, \prod_{j  \ = \  1}^\mathcal{N} G_{\,m}^{j}\,\bigl(\, \zeta^{j}\,\bigr) \right ]^2  } \,\right)
\end{equation}
\nomenclature[F]{$\varepsilon$}{Combined model error}

\subsection{Indicator for the CPU time}
A fair comparison of the computational time for various methods is not easy to undertake as it will depends on the way we code and the tools we use. To get a fair comparison, calculation times were measured on the same computer and on the same environment. Except for the reference solution calculation (where the \texttt{Matlab} toolbox \texttt{Chebfun} has been used), we developed all the other computational codes by ourselves. We paid attention to code each model in the same way (for example the same algorithm is always used to solve a system of equation). 
\bigbreak
For each step, the CPU calculation time is evaluated on a Lenovo, windows 10 with 8Go RAM IntelCore i5, 2.60 GHz. The CPU calculation time is normalized by the time constant $t_{0}$. It corresponds to the maximum CPU time observed. This information will be given in the titles of the figures. The CPU time ratio is noted $\rho_{\text{CPU}}$ and defined as follows: 
\begin{equation}
    \rho_{\text{CPU}} \ = \ \frac{t_{CPU}}{t_{0}}
\end{equation}

\nomenclature[F]{$\rho_{\text{CPU}}$}{Normalized CPU calculation time}
\nomenclature[F]{$t_{0}$ }{CPU time constant [$\,\mathsf{sec} $]}

\section{Theoretical case study}
\label{sec:theoretical_case_study}

\subsection{Description of the case study}
\subsubsection{Physical constants used}
\begin{figure}[htp]
    \begin{minipage}[c]{.46\textwidth}
  \centering
  \subfigure[Inside and outside air temperature signal]{\label{fig:u_o_i}\includegraphics[width=1\textwidth]{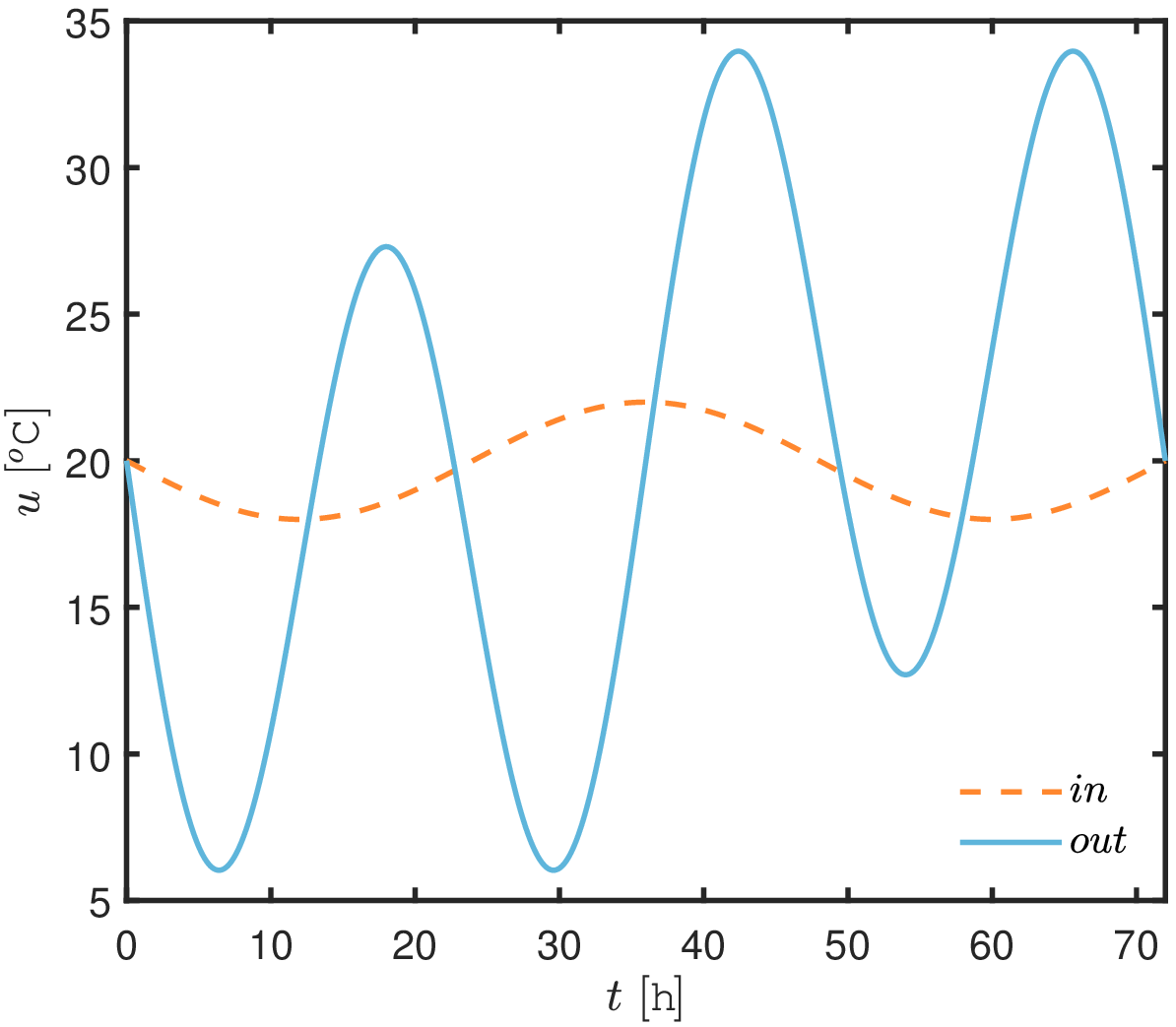}}
    \end{minipage}
    \hfill%
    \begin{minipage}[c]{.46\textwidth}
  \subfigure[Net radiative heat flux ]{\label{fig:qe}\includegraphics[width=1\textwidth]{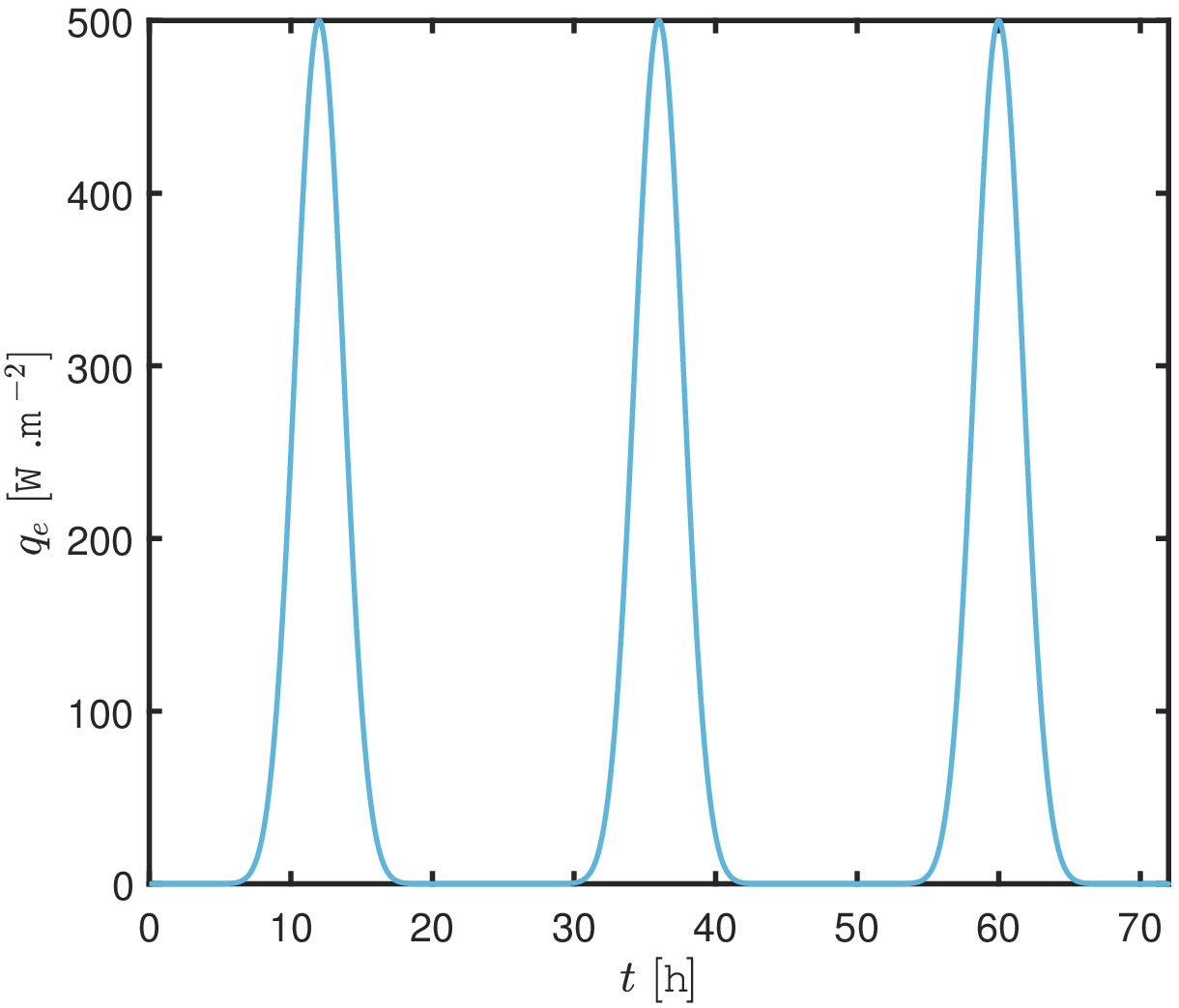}}
        \end{minipage}
  \caption{Boundary conditions of the theoretical case study}
  \label{fig:BC_Case_study}
\end{figure}

The case study consists of a wall of one-layer of thickness $\mathrm{L} \ = \ 0.20 \, \mathsf{m}$, made of concrete, with a thermal conductivity $\mathrm{k} \ = \ 1.75 \, \mathsf{W}. \mathsf{m}^{- 1}. \mathsf{K}^{-1}$ and a specific heat capacity $\mathrm{c} \ = \ 2.2 \, 10^{\,6} \,  \mathsf{J}. \mathsf{m}^{- 3}. \mathsf{K}^{-1}$. 
\bigbreak 
On the outdoor side of the wall, a sinusoidal variation of the air temperature and the net radiative heat flux are considered. Their variations are defined as: 
\begin{equation}
    \mathrm{u_{\,out} \ = \ u_{\, o, \, m} + \delta_{\, o,\, 1} \, \sin(\, 2 \ \pi \ \omega_{\, o,\, 1} \ t) + \delta_{\, o,\, 2} \, \sin(\, 2 \ \pi \ \omega_{\, o,\, 2} \ t)}
\end{equation}
\begin{equation}
    \mathrm{q \ = \ q_{ \, m} \,  \sin(\, 2 \ \pi \ \omega_{\, q,} \ t)^{\, 20}}
\end{equation}
On the indoor side of the wall, a sinusoidal variation of the air temperature is considered, as described below: 
\begin{equation}
    \mathrm{u_{\,in} \ = \ u_{\, i, \, m} + \delta_{\, i} \, \sin(\, 2 \ \pi \ \omega_{\, i, 1} \ t ) }
\end{equation}
As presented before, the net radiative heat flux is neglected on that side of the wall. The error due to this simplification of the mathematical model is studied for the specific case study in the appendix \ref{app:details_on_the_model_error}. 
\bigbreak 
The following numerical values are considered for the outdoor and indoor boundary conditions: 
\begin{equation*}
    \mathrm{u_{\, o, \, m}} \ = \ 20 \, [\, \degC \, ] , \; \; 
    \mathrm{\delta_{\, o,\, 1}} \ = \ - 4.4\, [\, \mathsf{K} \, ] , \; \; 
    \mathrm{\omega_{\, o,\, 1}}  \ = \ \frac{1}{ 72 }\, [\, \mathsf{h}^{-1} \, ], \; \; 
    \delta_{\, o,\, 2}  \ = \ - 11.7 \, [\, \mathsf{K} \, ], \; \;
    \omega_{\, o,\, 2} \ = \ \frac{1}{ 24 }\, [\, \mathsf{h}^{-1} \, ],
\end{equation*}
\begin{equation*}
    \mathrm{q_{ \, m}} \ = \ 500 \,[\, \mathsf{W}. \mathsf{m}^{- 2} \, ], \; \; 
    \omega_{\, q,} \ = \ \frac{1}{ 48 }\, [\, \mathsf{h}^{-1} \, ], \; \;
    \mathrm{u_{\, i, \, m}}  \ = \ 20\, [\, \degC \, ] , \; \; 
    \delta_{\, i}  \ = \ - 2.0\, [\, \mathsf{K} \, ], \; \;
    \omega_{\, i, 1} \ = \ \frac{1}{ 48 }\, [\, \mathsf{h}^{-1} \, ]. \; \;
\end{equation*}
Some of the numerical values are inspired from  1D numerical application \cite{berger2018intelligent}.  The boundary conditions used are presented in the Figure \ref{fig:BC_Case_study}. 
The convective heat transfer coefficients are set to : $\mathrm{h_{\, in}} \ = \ 8.7 \, \mathsf{W}. \mathsf{m}^{- 2}. K^{-1}$ and $\mathrm{h_{\, out}} \ = \ 23.2 \, \mathsf{W}. \mathsf{m}^{- 2}. \mathsf{K}^{-1}$. 
\bigbreak 

The numerical values of the dimensionless quantities are the following ones: 
\begin{equation*}
\, Bi_{\,in} : \ = \ 0.4971 ; \; \; \;
\, Bi_{\,out} : \ = \ 1.3314 ; \; \; \;
\, Fo : \ = \ 1 ; \; \; \;
\, \mathrm{t_{\,ref}} : \ = \ 1.2571 \times 10^{4} 
\end{equation*}

\subsubsection{Reference solution}

\begin{figure}[htp]
    \begin{minipage}[c]{.46\textwidth}
  \centering
  \subfigure[Indoor and outdoor surface temperature time series]{\label{fig:u_0}\includegraphics[width=1\textwidth]{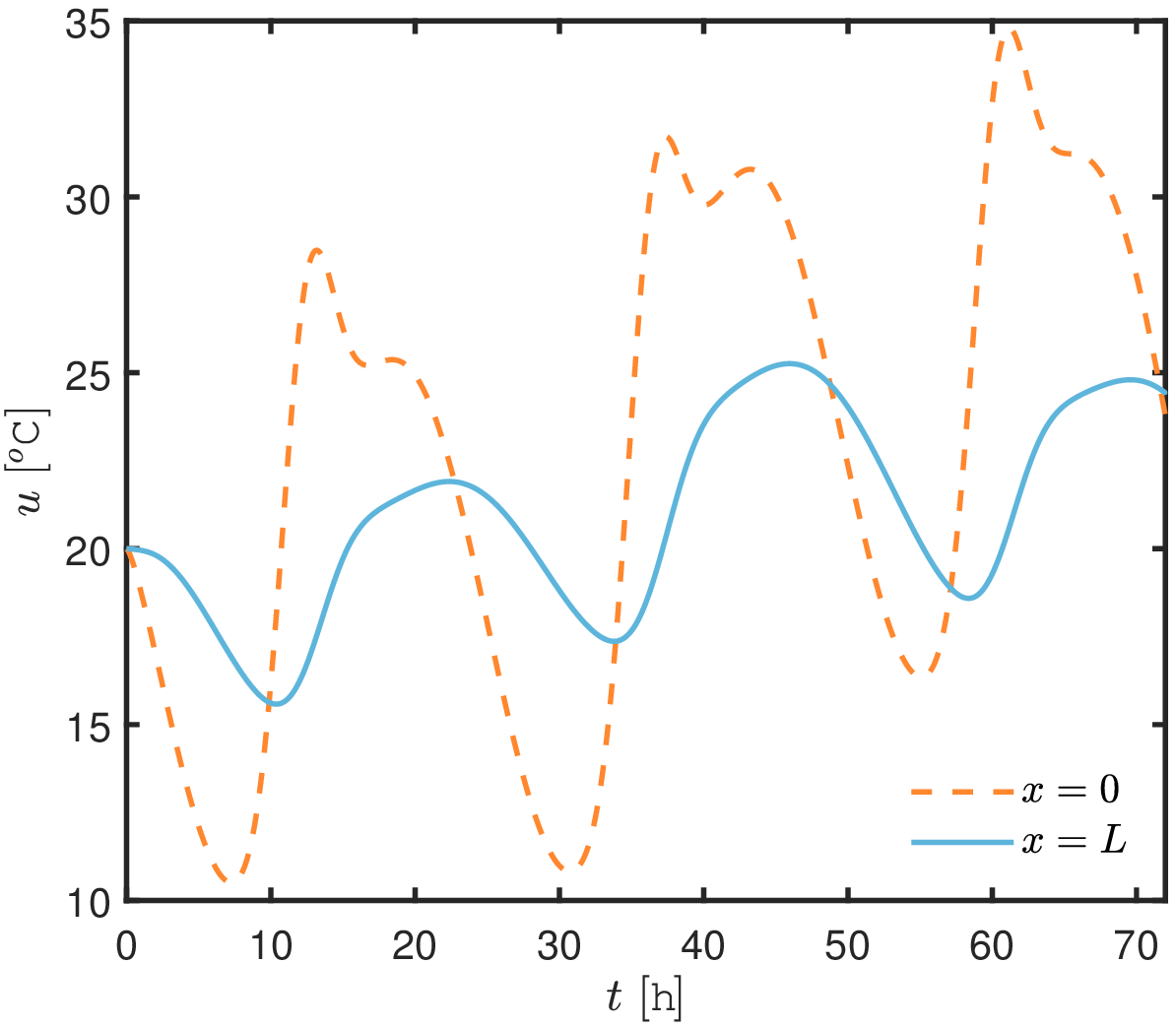}}
    \end{minipage}
    \hfill%
    \begin{minipage}[c]{.46\textwidth}
  \subfigure[Temperature profiles inside the wall ]{\label{fig:u_L}\includegraphics[width=1\textwidth]{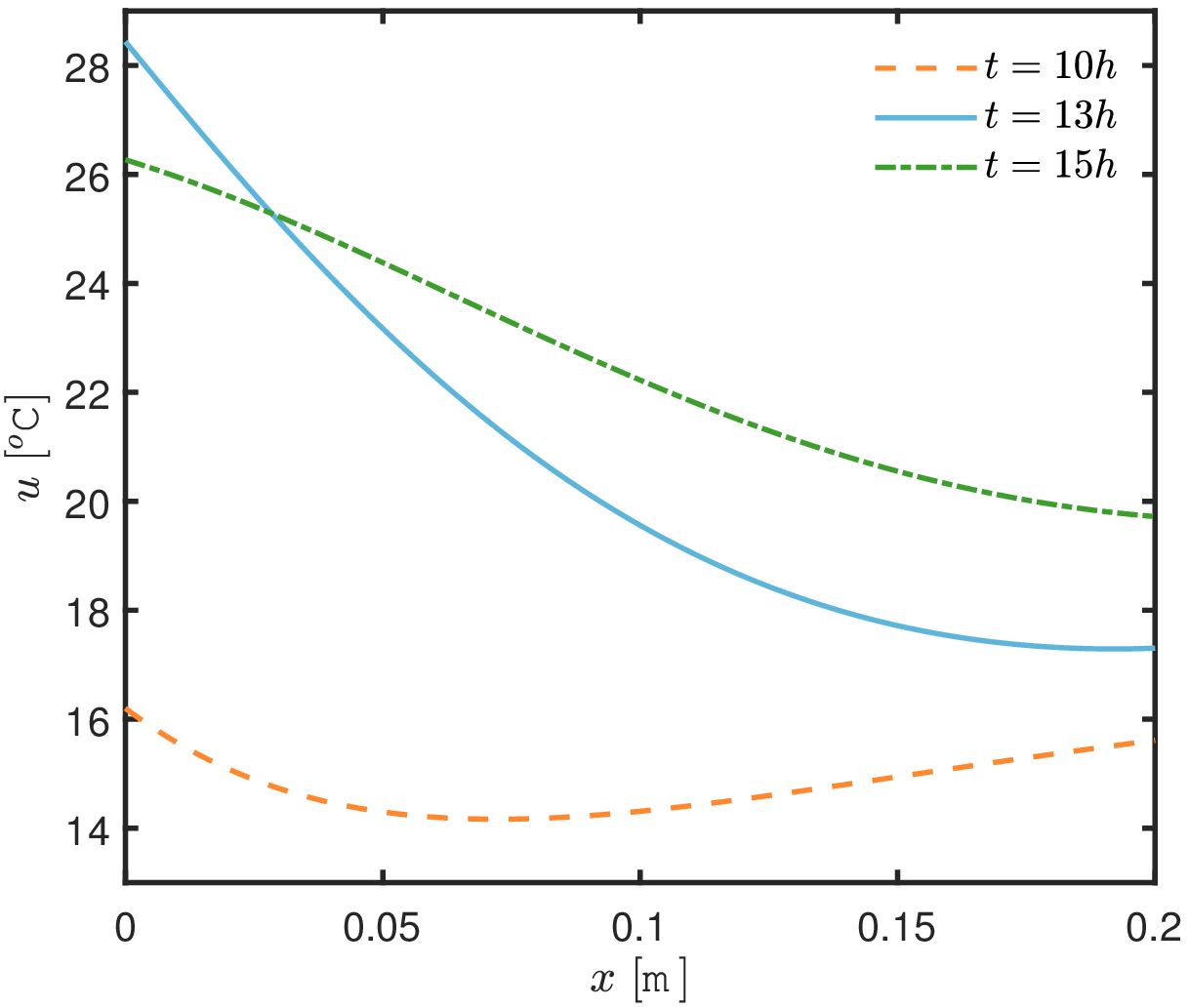}}
        \end{minipage}
  \caption{Temperature field for the reference solution}
  \label{fig:Sol_Ref_Case_study}
\end{figure}

The reference solution $y_{\, ref}(x,t)$ is computed using the \texttt{Matlab} toolbox \texttt{Chebfun} \cite{driscoll2014chebfun} for a time horizon of 3 days, with a dimensionless time step of  $\Delta t = 10^{-3}$  and a space mesh made of 200 nodes. The evolution of the temperature for the reference solution is presented in Figure \ref{fig:Sol_Ref_Case_study}. Figure \ref{fig:u_0} describes the temporal evolution of the surface temperature on each side of the wall and figure \ref{fig:u_L} gives an overview of the temperature profiles within the wall. It represents the source term that needs to be parameterized with the several studied approximation basis.

\subsubsection{Learning process}
As presented in the Section \ref{Sec:offline_online}, the POD basis is built on an available data-set. The choice was made to use the complete reference solution data-set to built the POD basis. The POD basis is then used in identical conditions than the one used for the learning process. Thus, the condition of the learning process will not influence the accuracy of the basis.

\subsection{Evaluation of the approximation of the source term}
\label{sec:4_2_source_term_theoretical_case}
The ability of each basis to approximate the source term depends on two parameters : the number of modes in the approximation basis $\mathcal{N}$ and the discretization of the parameters. The influence of those two criteria is studied hereafter. 

\subsubsection{Influence of the number of modes in the approximation basis}
\label{sec:4_2_1_inf_N_source_term}

\begin{figure}[htp]
    \begin{minipage}[c]{.46\textwidth}
  \centering
  \subfigure[ Semi-logarithmic scale ]{\label{fig:semilog}\includegraphics[width=1\textwidth]{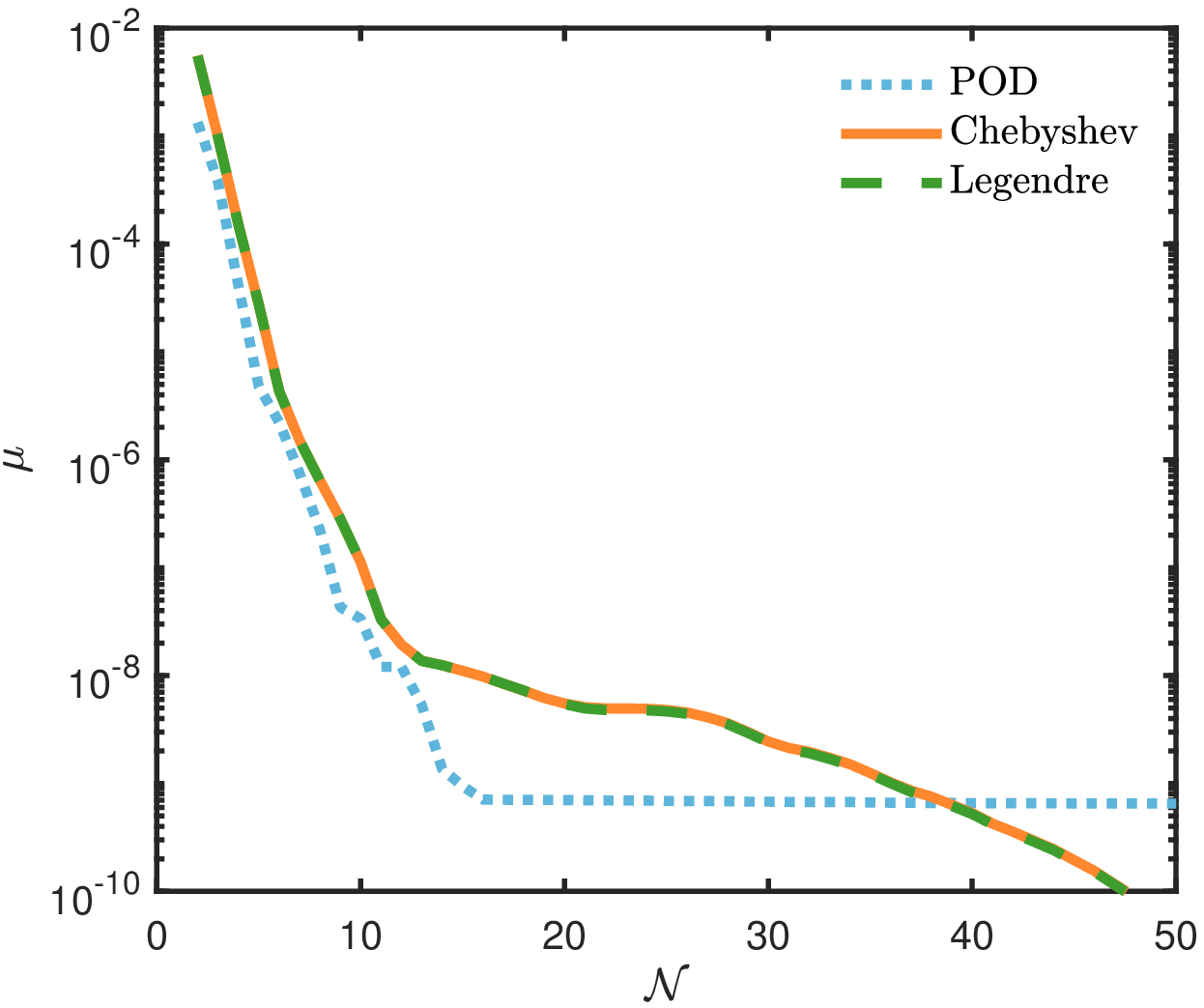}}
    \end{minipage}
    \hfill%
    \begin{minipage}[c]{.46\textwidth}
  \subfigure[ Logarithmic scale ]{\label{fig:loglog}\includegraphics[width=1\textwidth]{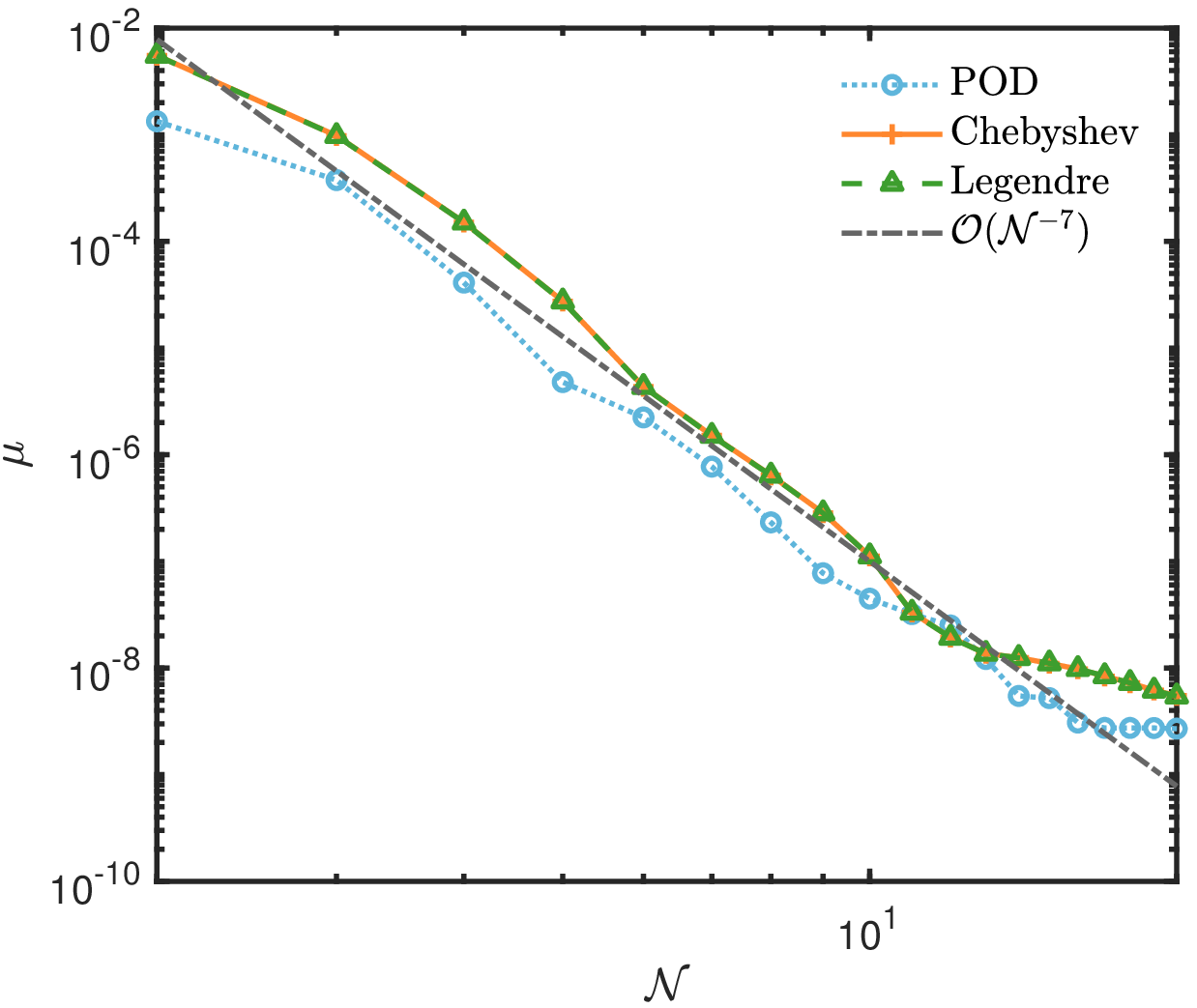}}
        \end{minipage}

  \caption{Evolution of the error $\mu$ as a function of the number of modes in the approximation basis}
  \label{fig:Comp_basis_Err}
\end{figure}

Figure \ref{fig:Comp_basis_Err} presents the evolution of the approximation error as a function of the number of modes $\mathcal{N}$ in the three approximation basis. In Figure \ref{fig:semilog}, we can observe that the error decreases as the number of modes in the approximation basis increases. In the case of the POD basis, the error decreases until it gets constant around $\mathcal{N}=18$. The results of the \textsc{Chebyshev} and the \textsc{Legendre} polynomial basis are very close. They both decrease with a large slope for the first ten modes and continue to decrease slowly. The polynomial basis cross the POD basis around $\mathcal{N}=38$ modes. The polynomial basis are then more accurate than the POD one.
\bigbreak 
The smoothness of the function can be linked to the number of times the function is differentiable. As explained by Trefethen (2013 \cite{trefethen2013approximation}), the smoother a function, the faster its approximates converge. Figure \ref{fig:loglog} gives information on the smoothness of the function. In this specific case, the three approximation basis have similar trends. They converge at a rate of $\mathcal{O}(\mathcal{N}^{-7})$.
\bigbreak 
\begin{figure} [htp]
    \centering
    \includegraphics[width=0.46\textwidth]{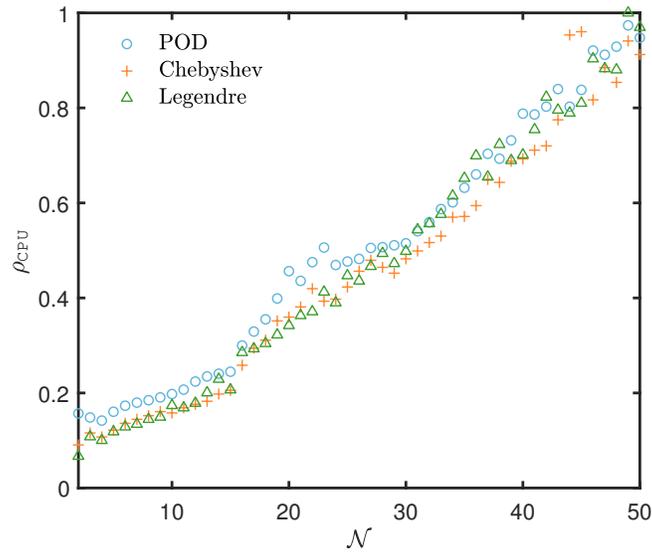}
    \caption{Evolution of the CPU calculation time as a function of the number of modes in the three approximation basis with $t_0 \ = \ 0.2341 \, \mathsf{sec}$}
    \label{fig:Comp_basis_CPU}
\end{figure}

The CPU calculation time is another criterion to compare the performance of the three basis. CPU time ratios are presented on Figure \ref{fig:Comp_basis_CPU}. The results are normalized by the maximum CPU  time observed (for \textsc{Legendre} basis with $\mathcal{N}=50$). The calculation time presented for the POD basis includes the learning process. We can observe that the CPU time increases linearly and that it is slightly higher for the POD basis than for the two polynomial basis. However, the results are of the same order of magnitude.

\subsubsection{Influence of the discretization}
\label{sec:4_2_2_inf_delta_source_term}

As reported in the Section \ref{sec:PGD}, each parameter of the model (the mesh, the boundary conditions, and the coefficients of the approximation basis) needs to be converted into vectors of parameters. For that purpose, their domain needs to be discretized, by converting the continuous functions into discrete values. The mesh of discretization of the parameter $\zeta$ has a direct impact on the accuracy of the approximation of the source term. The influence of the mesh of discretization $\Delta \overline{\zeta}$ is studied hereafter.
\bigbreak

\begin{figure}[htp]
    \begin{minipage}[c]{.46\textwidth}
  \centering
  \subfigure[ Evolution of the error for $\Delta \overline{\zeta} \ = \ 10^{-2}$ ]{\label{fig:Err_fin_Cri103}\includegraphics[width=1\textwidth]{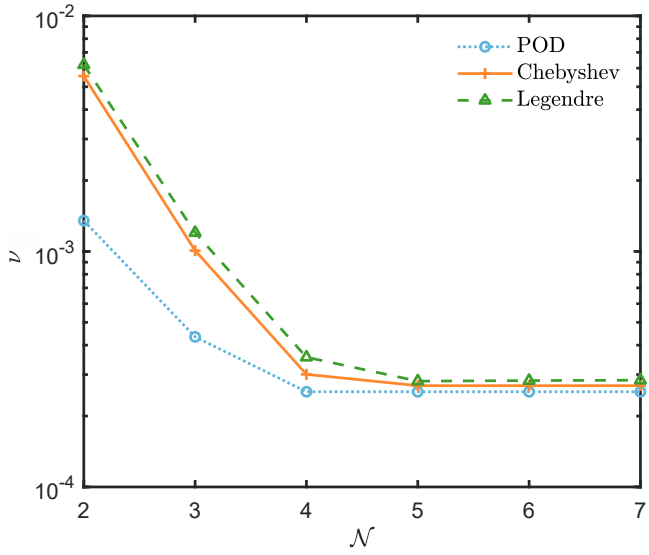}}
    \end{minipage}
    \vfill%
    \begin{minipage}[c]{.46\textwidth}
  \subfigure[ Evolution of the error for $\Delta \overline{\zeta} \ = \ 10^{-4}$ ]{\label{fig:Err_fin_Cri105}\includegraphics[width=1\textwidth]{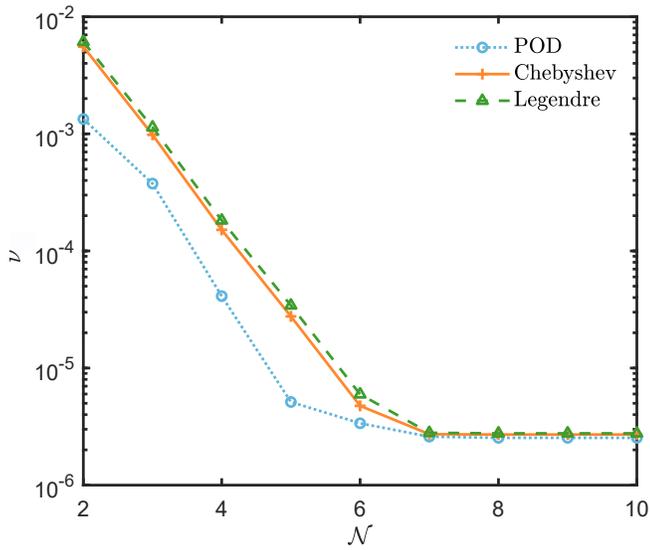}}
    \end{minipage}
    \hfill%
   \begin{minipage}[c]{.46\textwidth}
  \subfigure[ Evolution of the error for $\Delta \overline{\zeta} \ = \ 10^{-6}$ ]{\label{fig:Err_fin_Cri107}\includegraphics[width=1\textwidth]{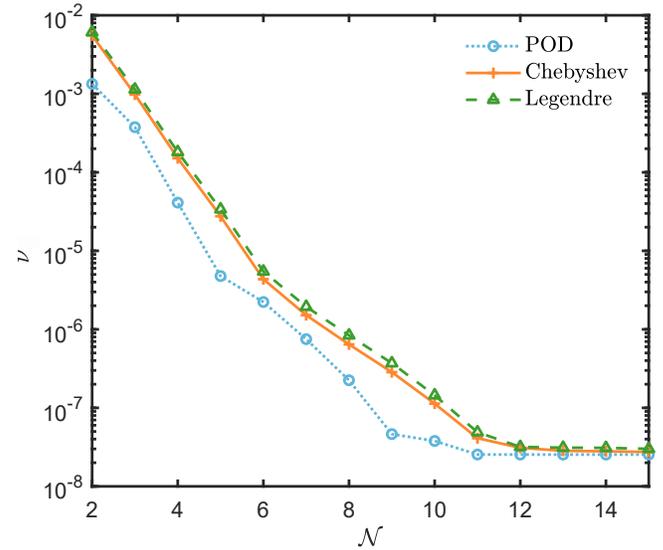}}
    \end{minipage}
  \caption{ Influence of the truncation of the approximation basis on the accuracy of the approximation for various discretizations}
  \label{fig:Err_fin_Cri}
\end{figure}
Three dimensionless discretizations have been selected $\Delta \overline{\zeta} \ = \ 10^{-2}, \, 10^{-4}, \, 10^{-6}$. For the three criteria, the error between the reference solution and the solution projected on the truncated basis is plotted as a function of the number of modes in the truncated basis. Figure \ref{fig:Err_fin_Cri} presents the results. For each curve, the same tendencies can be observed, the error drops and then stabilizes. Indeed as we increase the number of modes, the error of the approximation decreases. However, as the coefficients are rounded, part of the information is lost. As the error stabilizes, the addition of a supplementary mode does not improve the accuracy of the approximation. For a discretization  $\Delta \overline{\zeta} \ = \ 10^{-2}$, the threshold is reached for $\mathcal{N} \ = \ 5 $ and $7$ and $12$ modes for respectively: $\Delta \overline{\zeta} \ = \ 10^{-4}$ and $\Delta \overline{\zeta} \ = \ 10^{-6}$. 
\subsubsection{Discussion}
From those two first influence analyses, the approximation basis can not be ranked, as their performances are close. For both the number of modes $\mathcal{N}$ and the discretization, the same tendencies can be observed for the three approximation basis.
\bigbreak
Moreover, the accuracy of the POD basis depends on the quality of the learning process (it should be representative of the conditions of future study cases). In the theoretical case study, the learning process has been made on the complete reference solution data-set. We are then in ideal conditions for the use of the POD basis. In the practical application (Section \ref{sec:Practical application}), the influence of the learning process will be investigated.
\bigbreak
This first step enables the comparison of the behavior of the three studied basis outside of the PGD framework. However, once implemented in the PGD framework, the tendencies observed before could be different. to verify the consistency, the influence of the parameters studied should be studied in the PGD framework. 

\subsection{Evaluation of the PGD parametric model}
\label{sec: Combination of the approximation basis with the PGD parametric model}
A PGD parametric model is built to solve the problem studied here. The boundary conditions and the source term are defined as parameters of the parametric model. The approximation basis are used to describe the initial condition in a few parameters (modes). The PGD model is then combined to an approximation basis. The accuracy of the combined model depends on three parameters:

\begin{itemize}
    \item the accuracy of the approximation basis for a given number of modes $\mathcal{N}$,
    \item the discretization of each of the parameters vectors,
    \item the number of PGD modes $\mathcal{M}$.  
\end{itemize}

\bigbreak
To study the influence of those three parameters on the accuracy of the model, several PGD basis have been generated, one for each: combination of the three approximation basis, number of modes in the basis $\mathcal{N} \in [2, 5]$ and discretization $\Delta \overline{\zeta} \in [10^{-5}, 10^{-2}]$. In total 48 PGD combined models have been compared. For each model, both parameters of the alternating direction process and the enrichment process are fixed to $\tilde{\epsilon} = 10^{-6}$ and $\epsilon = 10^{-8}$. The influence of each parameter $\Delta \overline{\zeta,}$ $\mathcal{N}$, and $\mathcal{M}$ is studied hereafter, based on the results of those basis.

\subsubsection{Influence of the discretization of the approximation coefficient}

\begin{figure}[htp]
    \begin{minipage}[c]{.46\textwidth}
  \centering
  \subfigure[ $\varepsilon_\infty$ error for $\mathcal{N}=5$ ]{\label{fig:Err_dis_4m}\includegraphics[width=1\textwidth]{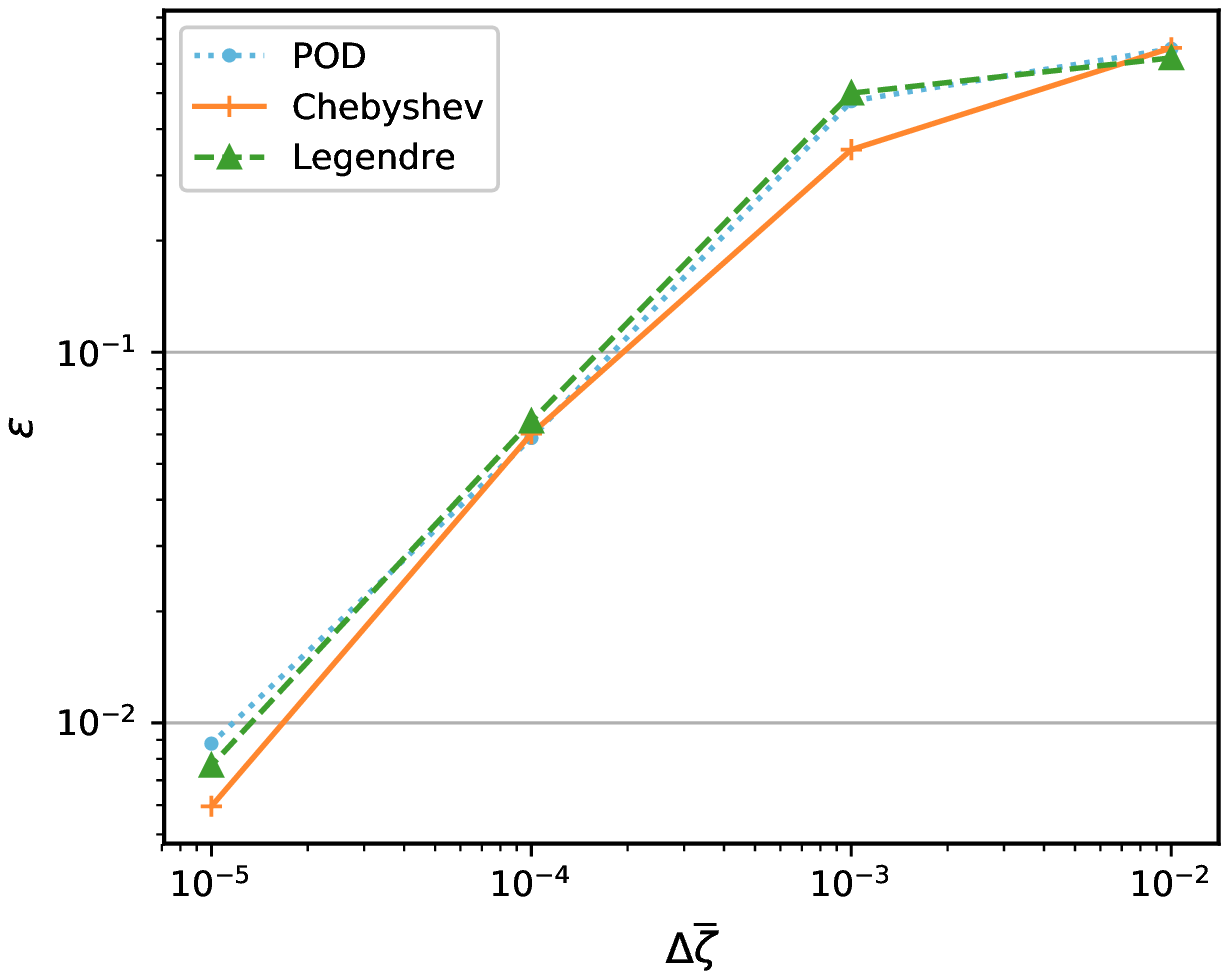}}
    \end{minipage}
        \hfill%
    \begin{minipage}[c]{.46\textwidth}
  \subfigure[Online CPU calculation time ratio for $\mathcal{N}=5$ ]{\label{fig:CPU_dis_4m}\includegraphics[width=1\textwidth]{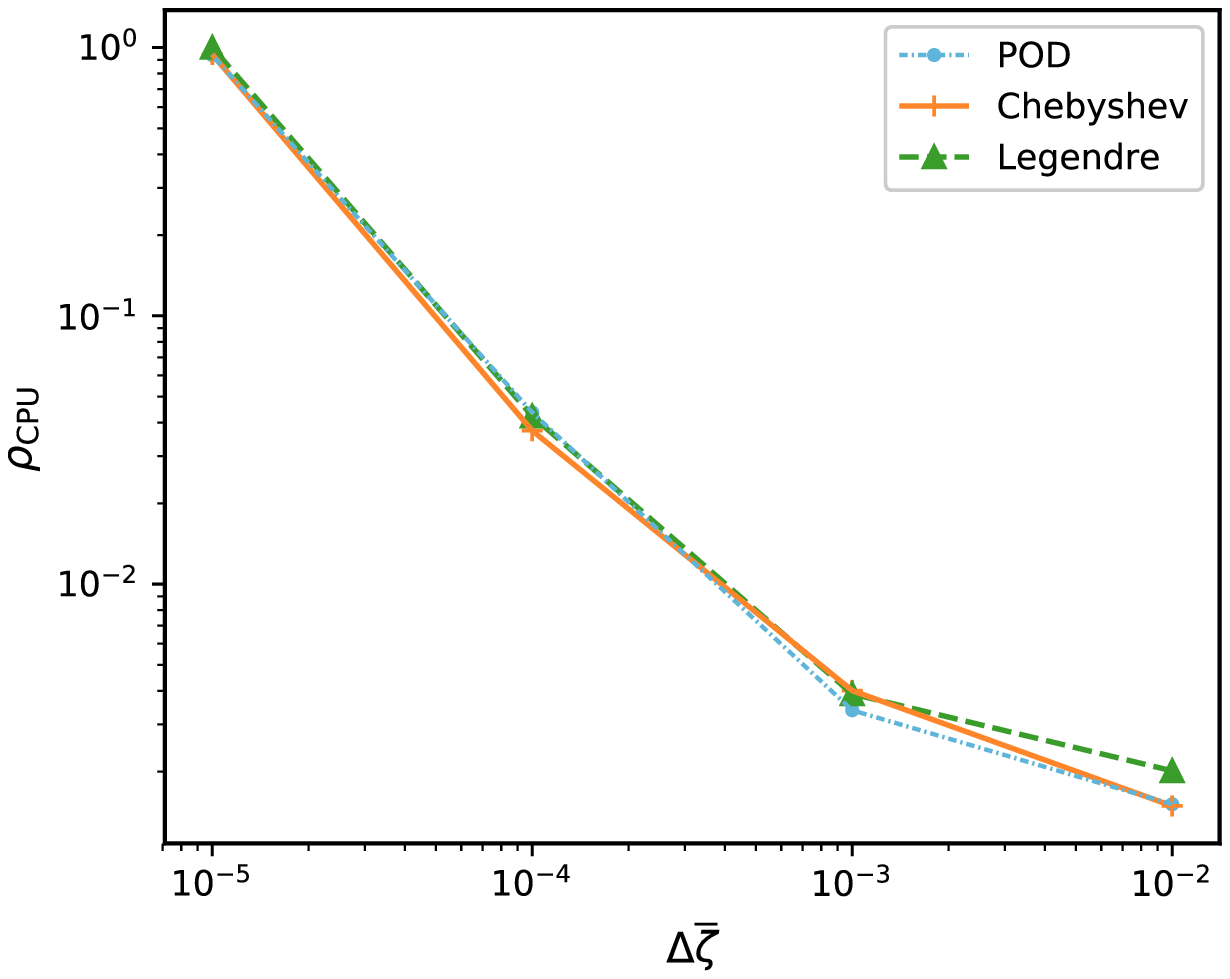}}
    \end{minipage}
  \caption{ Influence of the discretization of the approximation coefficients on the error and on the CPU time ratio with $t_0 \ = \ 23879 \, \mathsf{sec}$. }
  \label{fig:Err_dis_PGD}
\end{figure}

The influence of the approximation coefficients is studied here. Figure \ref{fig:Err_dis_PGD} presents the results of the error and about the calculation time. The results are displayed for the most accurate basis used made of $\mathcal{N}=5$ approximation modes. The accuracy of the model and the CPU time of each model increase, as the discretization gets finer. Those two results are in accordance with the previous ones. The discretization induces a loss of information. The continuous function is converted into discrete values as it is done for a spatial mesh for any numerical method. The finer the mesh, the closer the discrete representation to the continuous function. However, as we increase the discretization, we increase the number of elements in the vector. The online CPU time then increases. The same tendencies are observed for basis made of 3 and 4 modes. For basis made of 2 modes, the same tendencies are observed for the POD. However for the \textsc{Chebyshev} and \textsc{Legendre} basis, the error remains high and constant as we decrease the discretization. For both polynomial basis, using 2 modes is not enough to approximate the source term accurately.

\subsubsection{Influence of the number of modes in the approximation basis}
\label{sec:4_3_2_inf_N_PGD}

\begin{figure}[htp]
    \begin{minipage}[c]{.46\textwidth}
  \centering
  \subfigure[For $\Delta \overline{\zeta} = 10^{-5}$, evolution of the $\varepsilon$ error]{\label{fig:Err_N}\includegraphics[width=1\textwidth]{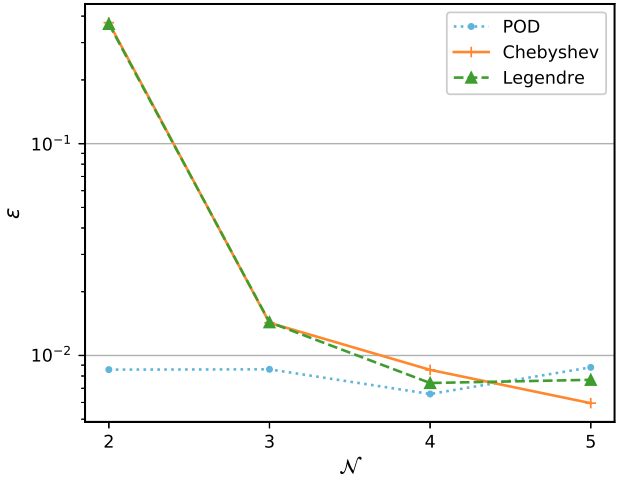}}
    \end{minipage}
       \hfill%
    \begin{minipage}[c]{.46\textwidth}
  \subfigure[For $\Delta \overline{\zeta}= 10^{-5}$, evolution of the CPU calculation time ratio ]{\label{fig:CPU_N}\includegraphics[width=1\textwidth]{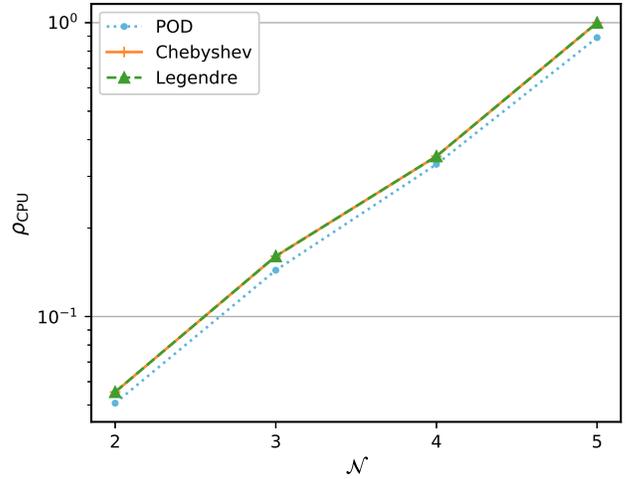}}
    \end{minipage}
  \caption{ Influence of the number of modes in the approximation basis on the $\varepsilon$ error and CPU time ratio with $t_0 \ = \ 23879 \, \mathsf{sec}$}
  \label{fig:Err_N_Basis}
\end{figure}

The influence of the number of modes in the approximation basis is now studied. Figure \ref{fig:Err_N} presents the evolution of the $\varepsilon$ error as a function of this parameter. Results are presented for a fixed discretization of $\Delta \overline{\zeta} \ = \ 10^{-5}$ for each approximation basis. In the case of the \textsc{Chebyshev} and \textsc{Legendre} combined parametric models, the error decreases with the number of modes. This phenomenon can be observed for fine discretizations ($\Delta \overline{\zeta} \ = \ 10^{-4}$ or $\Delta \overline{\zeta} \ = \ 10^{-5}$). For coarser discretizations, the error remains constant as we increase the number of modes. Adding a supplementary mode is not useful if the discretization remains constant.  
\bigbreak 
In the case of the POD combined parametric model, for a fixed discretization, adding a supplementary mode will not decrease the error of the model. The discretization will only have an impact on the error of the model. Here, the model is trained and used on the same data-set. The results may have been different if only part of the data-set has been used to train the basis. This point will be illustrated in the practical application (Section \ref{sec:Practical application}). 
\bigbreak 
For every model, a threshold around $\mathcal{O} (10^{-3}$) is reached after a few modes. The error of the final PGD model is then not mainly due to the approximation of the source term but also to other parameters: the discretization of the boundary condition on $\mathrm{x \ = \ L}$ fixed at $10^{-3}$, the discretization of the boundary condition on $\mathrm{x \ = \ 0}$ fixed at $10^{-4}$, the spatial grid fixed at $10^{-2}$.

\subsubsection{Influence of the number of modes in the PGD basis}

\begin{figure}[htp]
  \centering

  \includegraphics[width=.46\textwidth]{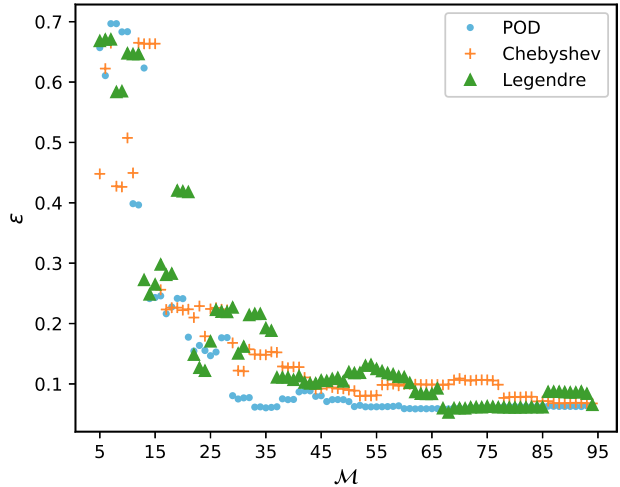}
  \caption{Influence of the number of PGD modes $\mathcal{M}$ on the $\varepsilon$ error for $\delta \overline{\zeta} = 10^{-4}$ and $\mathcal{N}=4$.}
  \label{fig:Err_M_Basis}
\end{figure}
The last parameter studied is the influence of the number of PGD modes $\mathcal{M}$. Figure \ref{fig:Err_M_Basis} presents the evolution of the error as a function of the number of PGD modes $\mathcal{M}$ for $\mathcal{N}=4$. The parametric models are built for $\Delta \overline{\zeta} \ = \ 10^{-4}$ for each approximation basis. This Figure gives information on how fast the PGD strategy converges. There are not many differences between the three methods. Applied to non-symmetric differential operators, the PGD algorithm converges slowly as its optimality is not guaranteed \cite{chinesta2013proper}. The PGD could contain more terms than strictly needed.
\bigbreak
\begin{figure}[htp]
    \centering
    \includegraphics[width=0.46\textwidth]{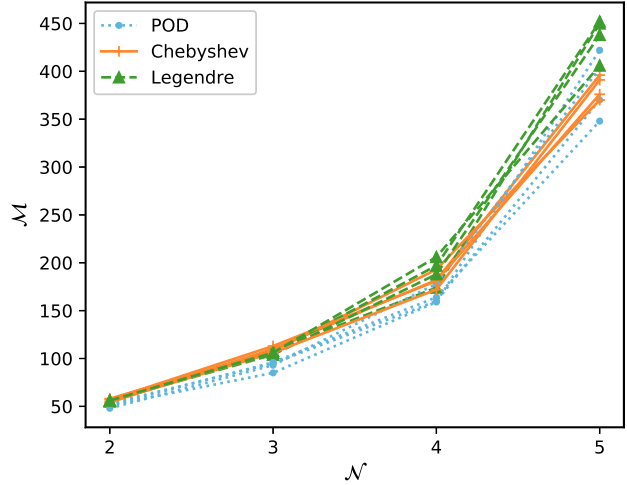}
    \caption{Evolution of the total number of PGD modes as a function of the number of modes of the approximation basis. Each curve of each basis corresponds to a different discretization $\Delta \overline{\zeta} \in [10^{-5}, 10^{-2}]$}
    \label{fig:N_f_M}
\end{figure}

\begin{figure}[htp]
    \begin{minipage}[c]{.46\textwidth}
  \centering
  \subfigure[For $\delta = 10^{-4}$, evolution of the $\varepsilon$ error]{\label{fig:Err_M_2}\includegraphics[width=1\textwidth]{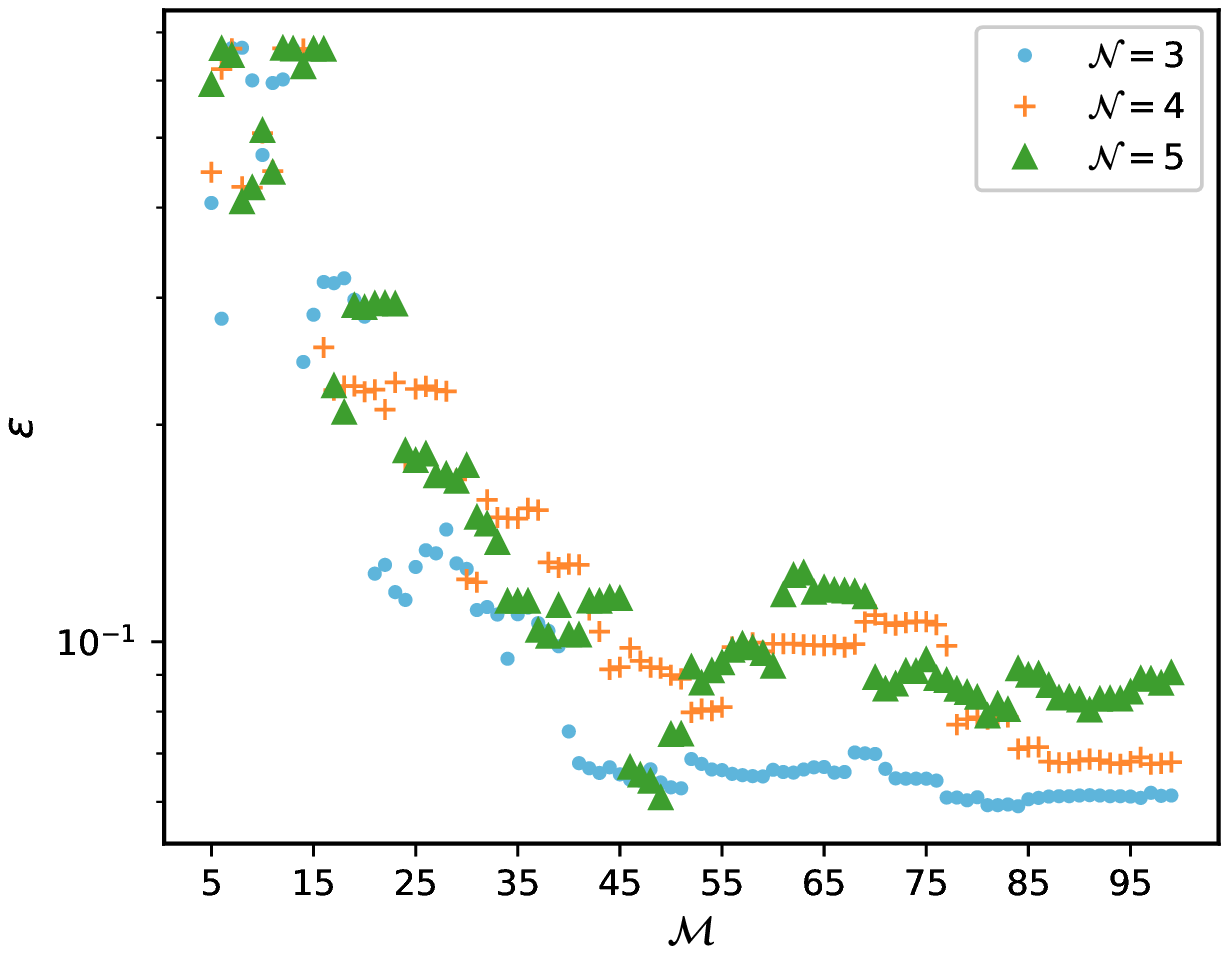}}
    \end{minipage}
        \hfill%
    \begin{minipage}[c]{.46\textwidth}
  \subfigure[For $\mathcal{N}=3$, evolution of the $\varepsilon$ error]{\label{fig:Err_M_3}\includegraphics[width=1\textwidth]{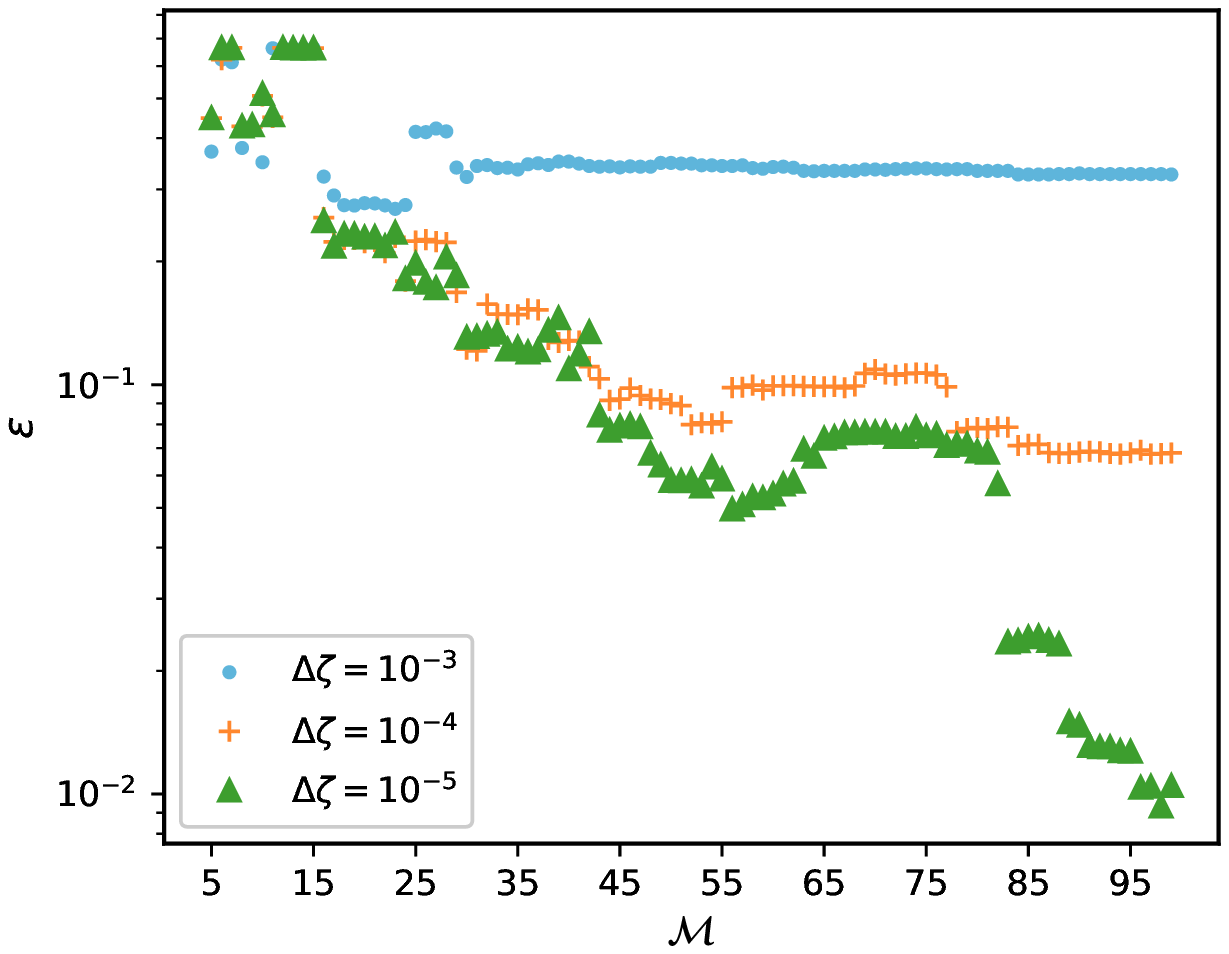}}
    \end{minipage}
  \caption{Evolution of the $\varepsilon$ error as a function of the number of modes for the \textsc{Chebyshev}  basis}
  \label{fig:Err_M_Basis_Cheb}
\end{figure}

Each time a mode is added to the parametric model, a new variable is added to the problem. The computational domain becomes of higher dimension, it must cover not only the physical and boundary conditions coordinates but also the parametric domain \cite{pruliere2010deterministic}. Adding a parameter increases the complexity of the tensor subspace. In the case of the PGD, this complexity grows linearly with the number of dimensions \cite{ammar2010error,borzacchiello2017non}. Figure \ref{fig:N_f_M} illustrates the impact of adding a new parameter to the PGD parametric model on the total number of PGD modes. As we increase the number of modes in the approximation basis, we increase the number of parameters in the parametric model. As soon in Figure  \ref{fig:Err_M_Basis_Cheb} the convergence rate of the algorithm decreases. Thus, the number of necessary PGD modes increases to achieve the desired accuracy ($\tilde{\epsilon} = 10^{-6}$ and $\epsilon = 10^{-8}$) as we increase the number of parameters involved.
\bigbreak 

\subsubsection{Discussion}
For the approximation coefficients discretization and the number of modes in the approximation basis, the same tendencies are observed than the one observed for the approximation of the source term. The comparison of the approximation basis on the approximation of the source term gives a good first overview of the behavior of the basis. 
\bigbreak
However, two modes are not sufficient to approximate the source term with \textsc{Chebyshev} and \textsc{Legendre} combined parametric models. For the POD basis, the final accuracy of the PGD model is reached with two modes for a fixed discretization. 
\bigbreak 
Finally, Leon \emph{et al.} (2018, \cite{leon2018wavelet}) have shown on the \textsc{Poisson} equation that the final accuracy of a PGD model depends on the discretization of the parameters and the number of terms $\mathcal{M}$ in the final sum. Indeed the finer the discretization of each parameter, the closer will be the discrete values to the continuous one. However, as they decrease the mesh, they increase the convergence rate of the PGD algorithm and the necessary number of PGD modes in the model. The same tendencies can be observed here in Figure \ref{fig:Err_M_3}. It presents the evolution of the $\varepsilon$ error as a function of the number of modes for the \textsc{Chebyshev} basis with $\mathcal{N}=3$. The error decreases and then reaches a threshold. Then adding a supplementary PGD mode to the parametric model is not sufficient to decrease the error of the model. The discretization should be decreased. 

\section{Practical application}
\label{sec:Practical application}
In the previous parts, the POD basis, as most of the time, has shown its optimality. However, as mentioned above, the performance of the POD basis depends on the quality of the learning process. It should be representative of the boundary conditions applied to the case study. 
\bigbreak
In the theoretical case study, the learning process has been made on the complete reference solution data-set. The POD basis is then used in identical conditions than the one used for the learning process. The influence of the learning process has not been studied yet. 
\bigbreak 
To obtain a POD basis, a training data-set is necessary. It can be obtained from measurements or from another numerical model. Both methods are expensive since a large range of data is needed. To give an example, if we want to use the parametric model to predict the temperature distribution in a wall during a year, the training data-set should be representative of all the boundary conditions encountered in practice.
\bigbreak
To illustrate this limit, the accuracy of various POD basis are compared to the polynomial basis. The same methodology as the one used for the theoretical case study is applied. The influence of the learning period is first studied on the approximation itself and then on the combination of the approximation basis with the PGD parametric model.
\bigbreak
Another major objective of this part is to evaluate the reliability of the model in a realistic case study. For that purpose, the results of the model are compared to laboratory measurements.

\subsection{Description of the case study}

\begin{figure} [htp]
    \begin{minipage}[c]{.46\textwidth}
    \centering
    \includegraphics[width=1.0\textwidth]{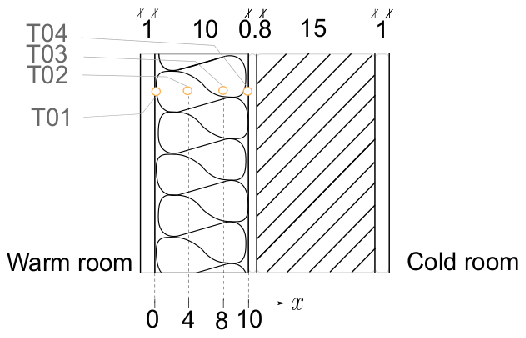}
    \caption{Sensors position illustration and nomenclature}
    \label{fig:Position_thermocouples}
    \end{minipage}
    \hfill%
    \begin{minipage}[c]{.46\textwidth}
    \centering
    \includegraphics[width=1.0\textwidth]{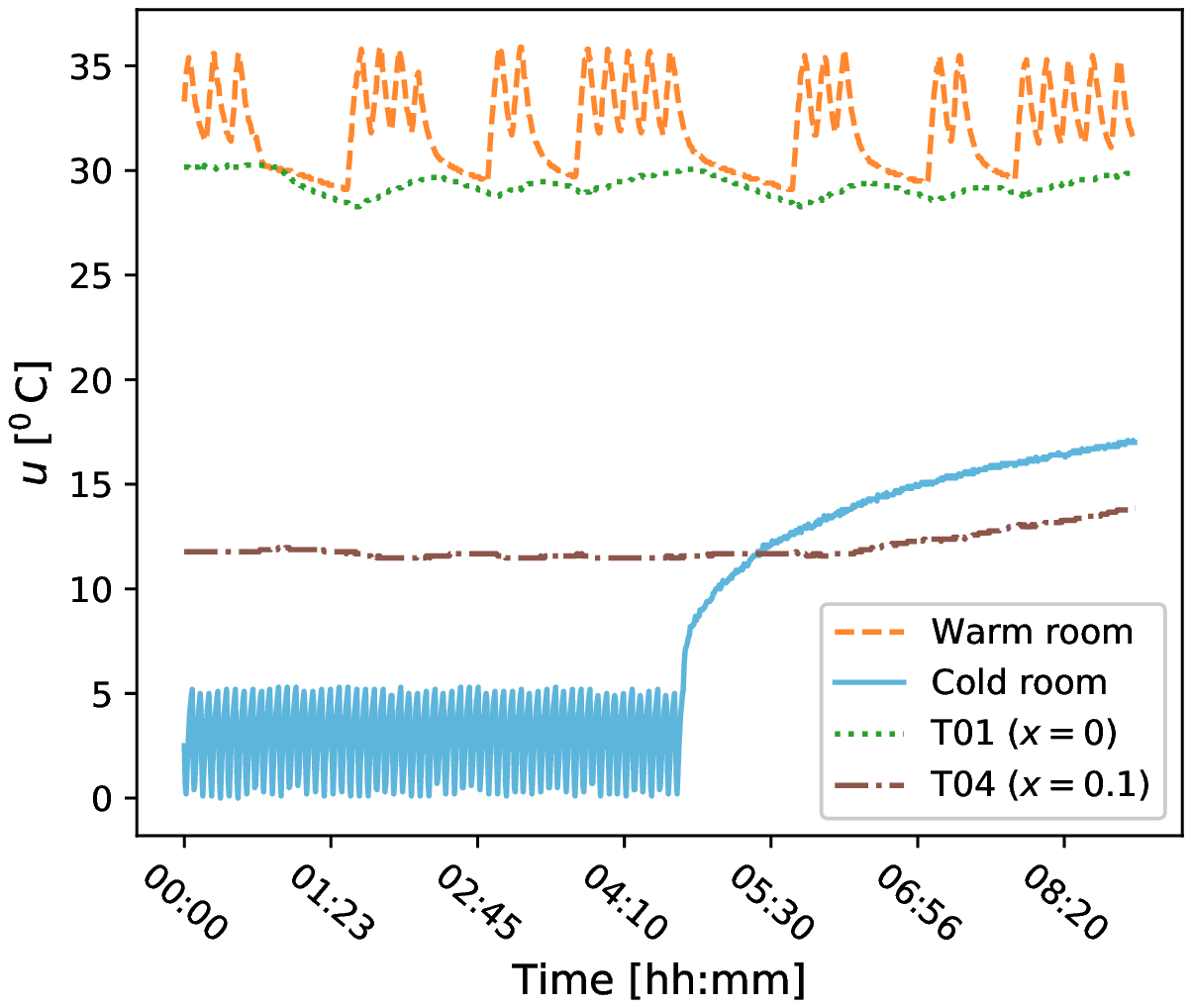}
    \caption{Evolution of the air temperature and inside wall temperatures measurements}
    \label{fig:Air_temp_Ch_Fr}
    \end{minipage}
\end{figure}



\subsubsection{Experimental set-up}

The experimental set-up described hereafter was think up with the objective to obtain realistic boundary conditions and measurements on a common building wall. It consists of a multi-layer building wall, made of traditional building materials: $1 \, \mathsf{cm}$ of plasterboard, $10 \, \mathsf{cm}$ of insulation (expanded polystyrene), $15 \, \mathsf{cm}$ of structural material and approximately $1 \, \mathsf{cm}$ of mineral coating. The wall is built between two rooms. One can be heated by an electric heater and the second one can be cooled by the evaporator of a heat pump.
\bigbreak 
For this study, we will only focus on the insulation layer of the wall. Indeed, the insulation material experiences greater temperature gradients which makes it more interesting to observe. Moreover, the insulation material is a homogeneous material and the temperature is easier to measure in such a material, contrary to the structure material made of concrete cellular blocks for which the measured temperature is strongly dependent on the position of the sensor. Indeed, for such cellular materials, the measured temperature can be very different whether it is measured on a cavity or near the wall of this cavity that creates a thermal bridge. By the more, the thermal conductivity of the insulation material is well known, whereas only the macroscopic thermal resistance is known for the concrete cellular block, which makes it difficult to obtain a calculated temperature directly comparable to the measured temperature, although the heat flux is correct. This insulation layer is thus equipped with four type K thermocouples located at the surface and in the insulation layer. 
\bigbreak
The global experimental uncertainty has been calculated with equation \ref{eq:uncertainty} \cite{taylor1997error}. 
\begin{equation}
    \sigma = \sqrt{ \sigma_{m}^{2} + \left( \frac{ \partial u}{ \partial x} \;  \delta x \right)^{2} }
    \label{eq:uncertainty}
\end{equation}
The thermocouples have been calibrated by measuring the temperature of melting ice and boiling water before the measurement. The sensor measurement uncertainty is then $\sigma_{m}= \pm 0.1^\circ \, \mathsf{C}$. The sensor position uncertainty has been evaluated as the product of the temperature derivative (with second-order centered approximation) at the sensor position and $\delta x = \pm 0.1 \, \mathsf{cm}$. The temporal mean global experimental uncertainty is noted hereafter $\overline{\sigma}$.

\subsubsection{Experimental observations}
Data were recorded for 5 days, with a $30 \, \mathsf{sec}$ time step. Several cycles were tested during this period, turning on and off the heater and/or the heat pump. The cycles are described in Table \ref{tab:boundary_conditions}. A pattern made of three cycles with three different time periods ($25 \, \mathsf{min}$, $40 \, \mathsf{min}$, and $60 \, \mathsf{min}$) is repeated twice. The first three cycles are run with a temperature set-point of $5 \, \mathsf{\degC}$ in the cold room. For the last three cycles, the heat pump was turned off to modify the boundary conditions of the cold room. The boundary conditions are described through the evolution of the air temperature in the warm and cold room in Figure \ref{fig:Air_temp_Ch_Fr}.

    \begin{table}[htp]
        \centering
        \begin{tabular}{|c|c|c|c|c|}
        \hline
          Cycle number  & Heater & Heat Pump & Duration [$\mathsf{min}$] & Time \\ \hline
          Initialization  & on & on ($5\, \mathsf{\degC} $) & 5220 ($87\mathsf{h}$) & - \\
          0  & off & on ($5\, \mathsf{\degC} $) & 40 & 00:00 to 00:40 \\
          1  & on & on ($5\, \mathsf{\degC} $) & 40 &  00:40 to 1:20 \\
          1  & off & on ($5\, \mathsf{\degC} $) & 40 &  1:20 to 2:00 \\
          2  & on  & on ($5\, \mathsf{\degC} $) & 25 &  2:00 to 2:25 \\
          2  & off & on ($5\, \mathsf{\degC} $) & 25 & 2:25 to 2:50 \\
          3  & on  & on ($5\, \mathsf{\degC} $) & 60 &  2:50 to 3:50  \\
          3  & off & on ($5\, \mathsf{\degC} $) & 60 &  3:50 to 4:50 \\
          4  & on & off & 40 &  4:50 to 5:30 \\
          4  & off & off & 40 &  5:30 to 6:10 \\
          5  & on & off & 25 &  6:10 to 6:35  \\
          5  & off & off & 25 &  6:35 to 7:00\\
          6  & on & off & 60 &  7:00 to 8:00  \\
          6  & off & off & 60 &  8:00 to 9:00\\
          \hline
        \end{tabular}
        \caption{Description of the cycles}
        \label{tab:boundary_conditions}
    \end{table}

\subsubsection{Reference solution}

The reference solution $y_{\, ref}(x,t)$ of this problem is computed using a Euler implicit finite difference model for a time horizon of 96 hours with a time step of $30 \mathsf{sec}$ (dimensionless time step of $\Delta = 10^{-2}$) and a spacial mesh made of 99 nodes. On each side, two \textsc{Dirichlet} boundary conditions are set using the temperature signal $T01$ and $T04$. Figure \ref{fig:Air_temp_Ch_Fr} gives an overview of the temperature evolution at the boundary conditions.
\bigbreak
\begin{figure} [htp]
    \centering
    \includegraphics[width=0.46\textwidth]{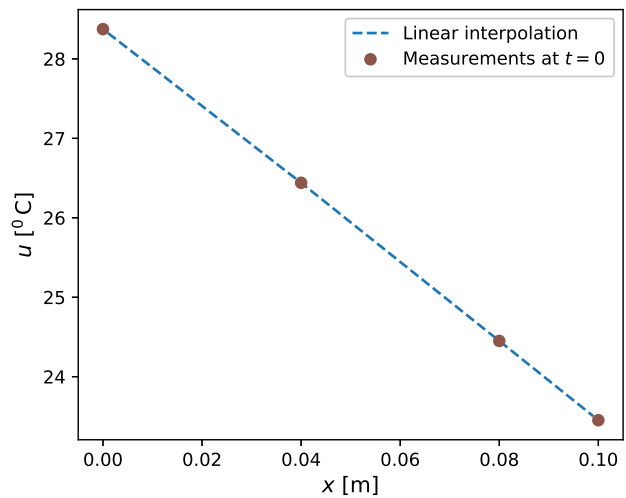}
    \caption{Initialization temperature profile}
    \label{fig:Initialization}
\end{figure}

The first temperature profile is initialized using the temperature profile measured in the wall at the beginning of the experiment. Linear interpolation is done between the measured points to obtain the temperature distribution at each point of the spatial mesh (Figure \ref{fig:Initialization}). The simulation is then run for the all period (5 days). The first $87\mathsf{h}$ of the simulation are not used. They are left as initialization period of the model. It consists of turning on the heater and the heat pump until an equilibrium between the two rooms is reached. The boundary conditions of this initialization cycle are described in Table \ref{tab:boundary_conditions}. The rest of the data-set is used to evaluate the model in different conditions. As the first $87\mathsf{h}$ are not used to evaluate the model, they are not presented in the following figures. Thermal properties from the French regulations database \cite{RT-2012} are used with: $k = 0.04 \,\mathsf{W}.\mathsf{m}^{-1}.\mathsf{K}^{-1}$ and $c = 30.10^{3} \,\mathsf{J}.\mathsf{m}^{-3}.\mathsf{K}^{-1}\,$. 

\subsubsection{Learning process}

Three training data-sets for the POD basis are compared:

\begin{enumerate}
    \item the full evaluation data-set (noted $t \in \Omega_{\tau} = [ 0 \, , \, \tau] $ with $\tau = 9 \mathsf{h}$),
    \item half of the evaluation data-set, made of the cycles 0 to 3 (noted $t \in \Omega_{\frac{\mathtt{\tau}}{2}} = [ 0 \, , \, 4\mathsf{h}50] $),
    \item the cycle 1 (noted $t \in \Omega_{t_{1}} = [ 0 \, , \, 0\mathsf{h}40] $).
\end{enumerate}

The three basis are compared to the \textsc{Chebyshev} and \textsc{Legendre} polynomial basis. For those two last methods, no learning period is required to build the basis.

\subsection{Influence of the learning period}
As for the previous case study, the influence of the learning period is first evaluated on the approximation of the source term, then it is evaluated on the PGD parametric model. The same parameters and computation code than the one used for the previous sections are applied. 

\subsubsection{Evaluation of the approximation of the source term}

\begin{figure}[htp]
    \centering
    \includegraphics[width=0.46\textwidth]{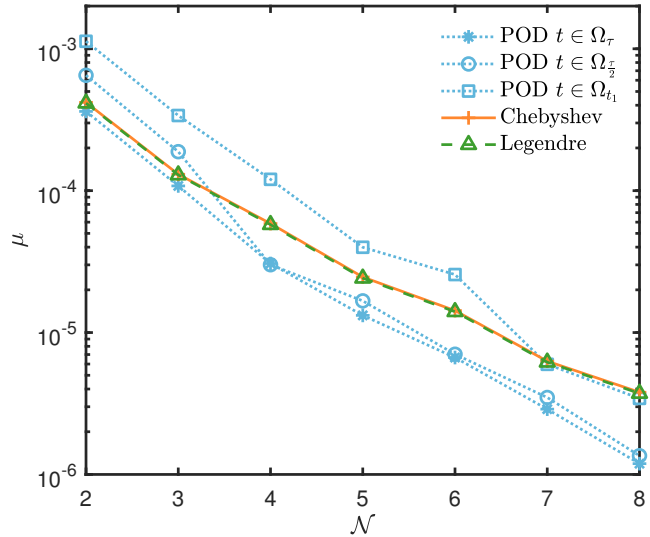}
    \caption{Approximation basis error as a function of the number of modes for several training periods}
    \label{fig:inf_learning period}
\end{figure}

The accuracy of the approximation of the source term is studied for various training data-sets. The results are presented on figure \ref{fig:inf_learning period}. The error is plotted for the various number of modes in the approximation basis: $\mathcal{N} \in [2,8]$. 

\bigbreak 
As in the theoretical example, the POD basis is the most accurate one for $\mathcal{N} \in [2, 8]$, if the full data-set is used for the training period. However if only a part of the data is available, the \textsc{Chebyshev} and \textsc{Legendre} approximation basis are more efficient for $\mathcal{N} \in [2, 3]$. The POD basis trained with half of the cycles seems to be as efficient as the one built with the full training data-set for $\mathcal{N} \in [4, 8]$. Indeed the same pattern is repeated from cycle 1-2-3 to cycle 4-5-6. Building the POD basis with one pattern could be enough for $\mathcal{N} > 3$.
\bigbreak 
This learning process has a numerical cost as it requires running a large original model and building the POD basis as described in Section \ref{Sec:offline_online}. Table \ref{tab:offline_calcul_time} compares the computation time needed to build the basis from the results of the finite difference model for the various learning periods. As large is the training data-set as large is the time needed. Building the basis with one cycle results in a saving of $35 \%$ of the offline computation cost. 
    \begin{table}[htp]
        \centering
        \begin{tabular}{|c|c|c|}
        \hline
          Learning period  & $\rho_{CPU}$ \\ \hline
          $t \in \Omega_{\tau}$  & 1 \\
          $ t \in \Omega_{\frac{\mathtt{\tau}}{2}}$  & 0.89  \\
          $ t \in \Omega_{t_{1}}$  & 0.65 \\
          \hline
        \end{tabular}
        \caption{Offline calculation time $t_0 = 0.003987\, \mathsf{sec}$}
        \label{tab:offline_calcul_time}
    \end{table}

\bigbreak 
Finally, a compromise should be found to minimize the training period and the computational cost needed to build the basis while keeping an accurate approximation basis. For that purpose, a methodology to select an efficient training period should be developed.

\subsubsection{Evaluation of the PGD parametric model}

The influence of the learning period is now studied for the combination of the PGD parametric model with the various approximation basis. Results for the most favourable ($t \in \Omega_{\tau} $) and unfavourable ($t \in \Omega_{t_1} $) POD basis are compared to the \textsc{Chebyshev} and \textsc{Legendre} polynomial basis. Several PGD basis have been generated one for each: combination of the four approximation basis (the favourable POD, the unfavourable POD, the \textsc{Chebyshev} and \textsc{Legendre} polynomial basis), number of modes $\mathcal{N} \in [2,5]$ and discretization $\Delta \overline{\zeta} \in [10^{-5}, 10^{-4}]$. In total 32 PGD basis have been compared for this application. As done before, both parameters of the alternating direction process and the enrichment process are fixed to $\tilde{\epsilon} = 10^{-6}$ and $\epsilon = 10^{-8}$. The accuracy of the PGD parametric model is compared for various number of modes $\mathcal{N} \in [2,5]$ and for a fixed discretization $\Delta \overline{\zeta} = 10^{-5}$. We use the same parameters than in section \ref{sec:4_3_2_inf_N_PGD} to compare the results. 
\bigbreak 
\begin{figure}[htp]
    \begin{minipage}[c]{.46\textwidth}
  \centering
  \subfigure[For $\Delta \overline{\zeta} = 10^{-5}$, evolution of the $\varepsilon$ error]{\label{fig:Err_N_learning_period}\includegraphics[width=1\textwidth]{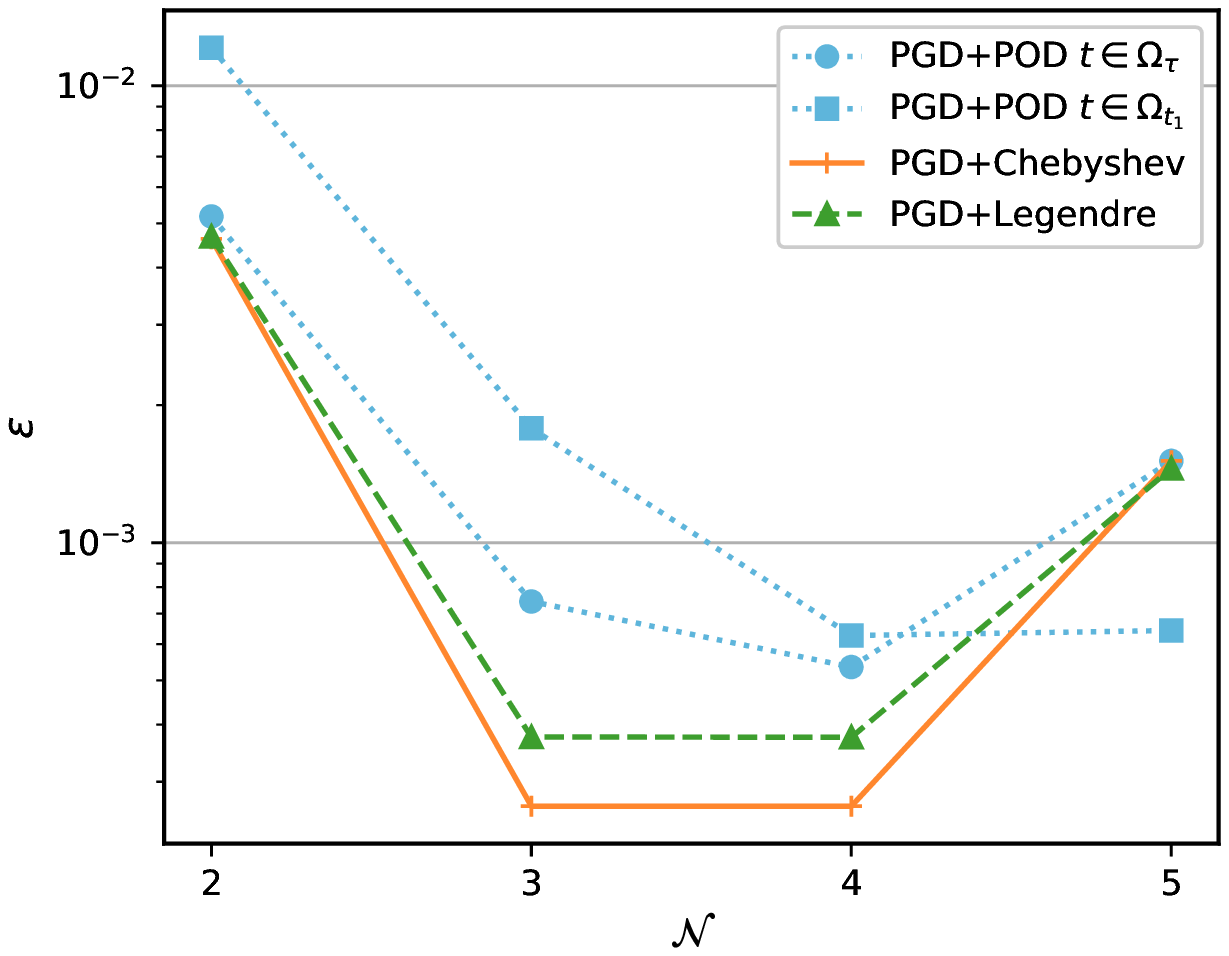}}
    \end{minipage}
        \hfill%
    \begin{minipage}[c]{.46\textwidth}
  \subfigure[For $\Delta \overline{\zeta}= 10^{-5}$, evolution of the CPU time ratio ]{\label{fig:CPU_N_learning_period}\includegraphics[width=1\textwidth]{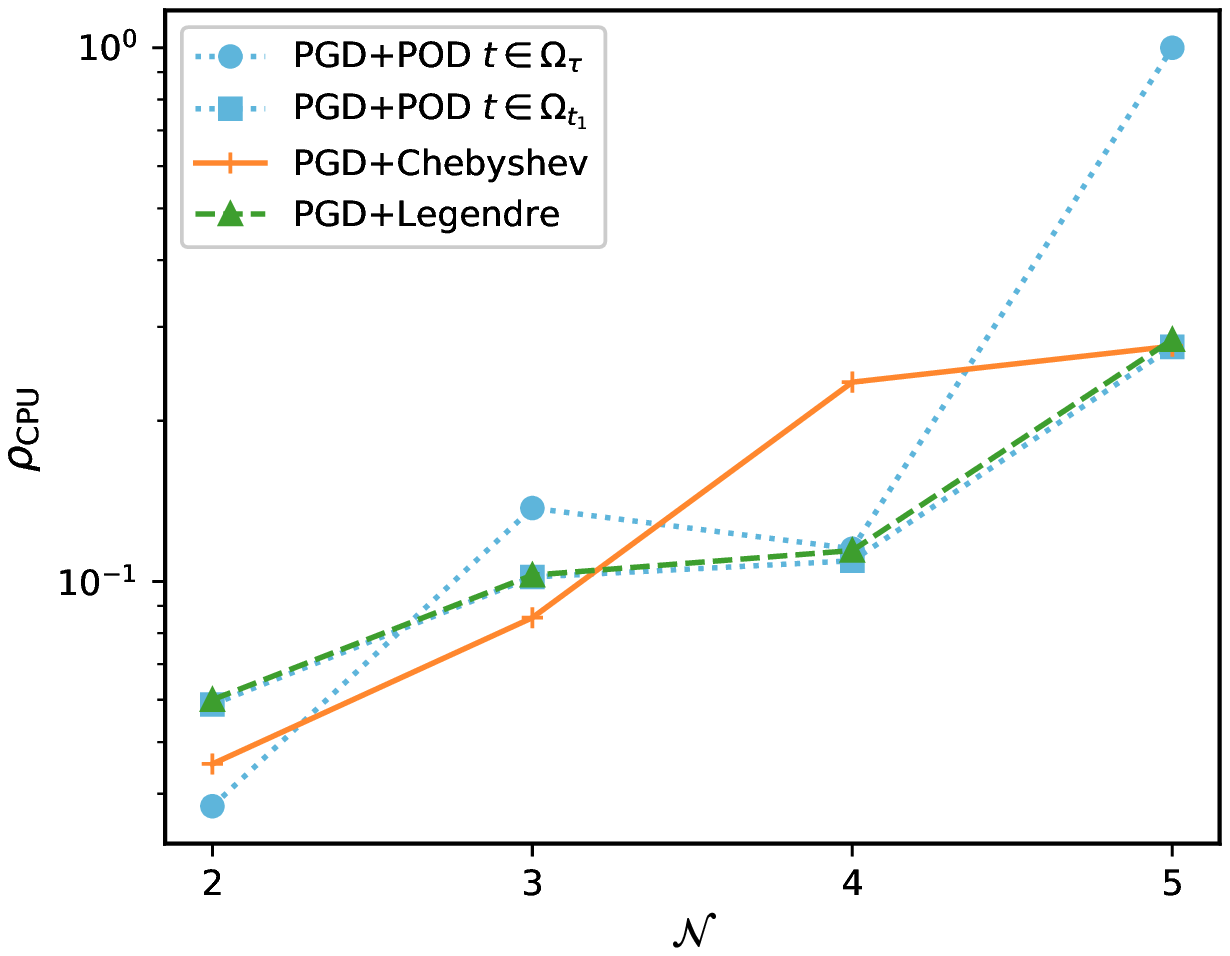}}
    \end{minipage}
  \caption{PGD parametric model $\varepsilon$ error and CPU time ratio as a function of the number of modes for several training periods ( $t_0 \ = \ 51.31 \, \mathsf{sec}$) }
  \label{fig:inf_learning period_PGD}
\end{figure}

Figure \ref{fig:inf_learning period_PGD} presents the evolution of the error and CPU time for various numbers of modes $\mathcal{N}$. For each combined parametric model, for $\mathcal{N} \in [2,4]$ the error decreases with the number of modes. The error for $\mathcal{N} = 5$ increases. This can also be observed in Figure \ref{fig:Err_N_Basis}. As previously explained, a threshold (around $\mathcal{O} (10^{-3})$) is reached after a few modes. This phenomenon can be observed for both discretizations. One this threshold has been reached, the error of the final PGD model is then not mainly due to the approximation of the source term. This could explain the fact that the error slightly increases. 
\bigbreak 
In the theoretical case study, the error of the POD basis remained constant with the number of modes. Adding supplementary modes did not improve the total accuracy of the model. It is not the case here. For more complex boundary conditions (realistic signal), supplementary modes are necessary to accurately parametrize the previous temperature profile. 
\bigbreak 
For this practical example, the \textsc{Chebyshev} and \textsc{Legendre} polynomial basis are more accurate once combined with the PGD basis for a similar computational time. This could be due, once more, to the complexity of the boundary condition signal. It could be also due to the discretization $\Delta \overline{\zeta}$. To encounter the same method ranking as the one presented in Figure \ref{fig:inf_learning period}, the POD coefficients may need to be discretized more finely.
\bigbreak 

\subsection{Comparison with experimental data}
\begin{figure}[htp]
   \begin{minipage}[c]{1.\textwidth}
  \centering
  \subfigure[Temperature at $4 \mathsf{cm} $ from the inner boundary condition]{\label{fig:Th_03}\includegraphics[width=1\textwidth]{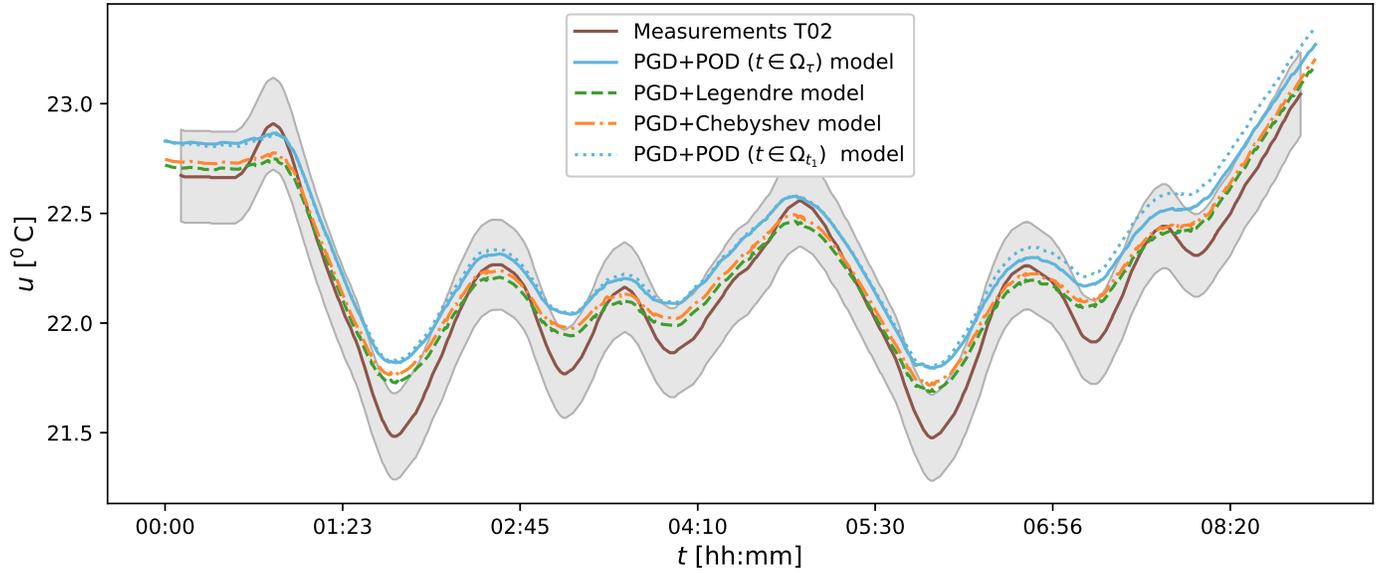}}
    \end{minipage}
        \vfill%
    \begin{minipage}[c]{1.\textwidth}
  \subfigure[ Temperature at $8 \mathsf{cm} $ from the inner boundary condition ]{\label{fig:Th_04}\includegraphics[width=1\textwidth]{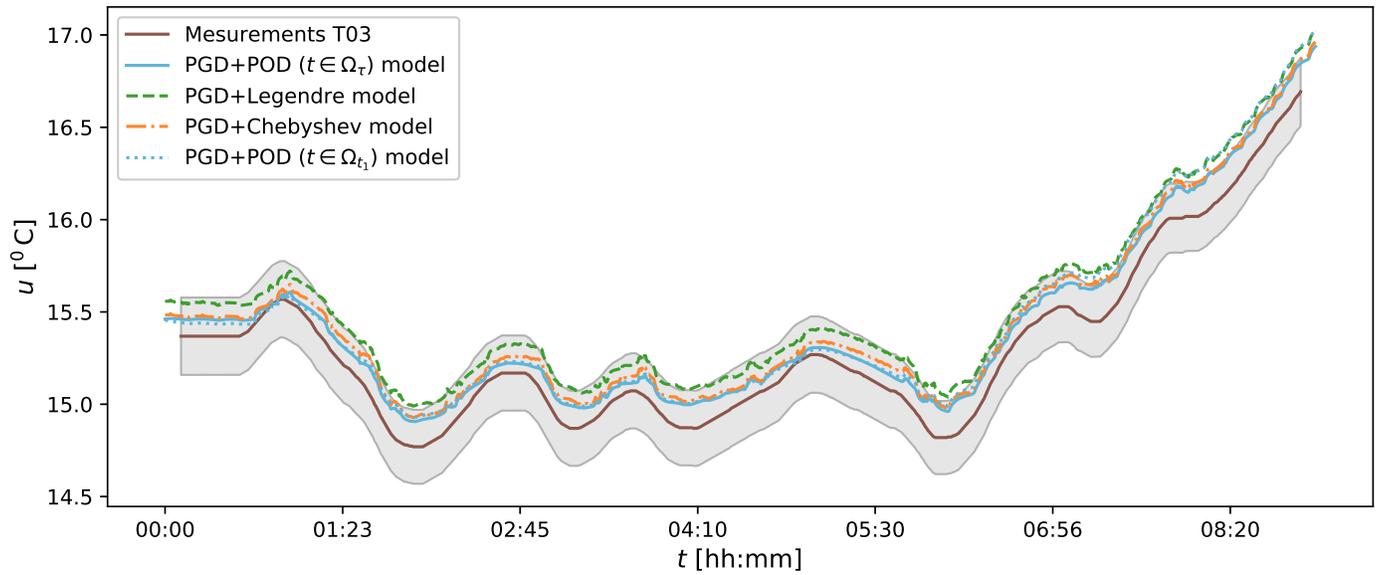}}
    \end{minipage}
  \caption{Time evolution of the temperature measured and calculated by the models at various depths. The grey zone corresponds to $\pm \sigma$, the global experimental uncertainty.}
  \label{fig:Comp_expe_model}
\end{figure}

\begin{figure}[htp]
    \begin{minipage}[c]{.46\textwidth}
  \centering
  \subfigure[For $\Delta \overline{\zeta} = 10^{-5}$, evolution of the $\varepsilon$ error at $4 \mathsf{cm}$ from the inner boundary condition]{\label{fig:Err_N_learning_period_Th03}\includegraphics[width=1\textwidth]{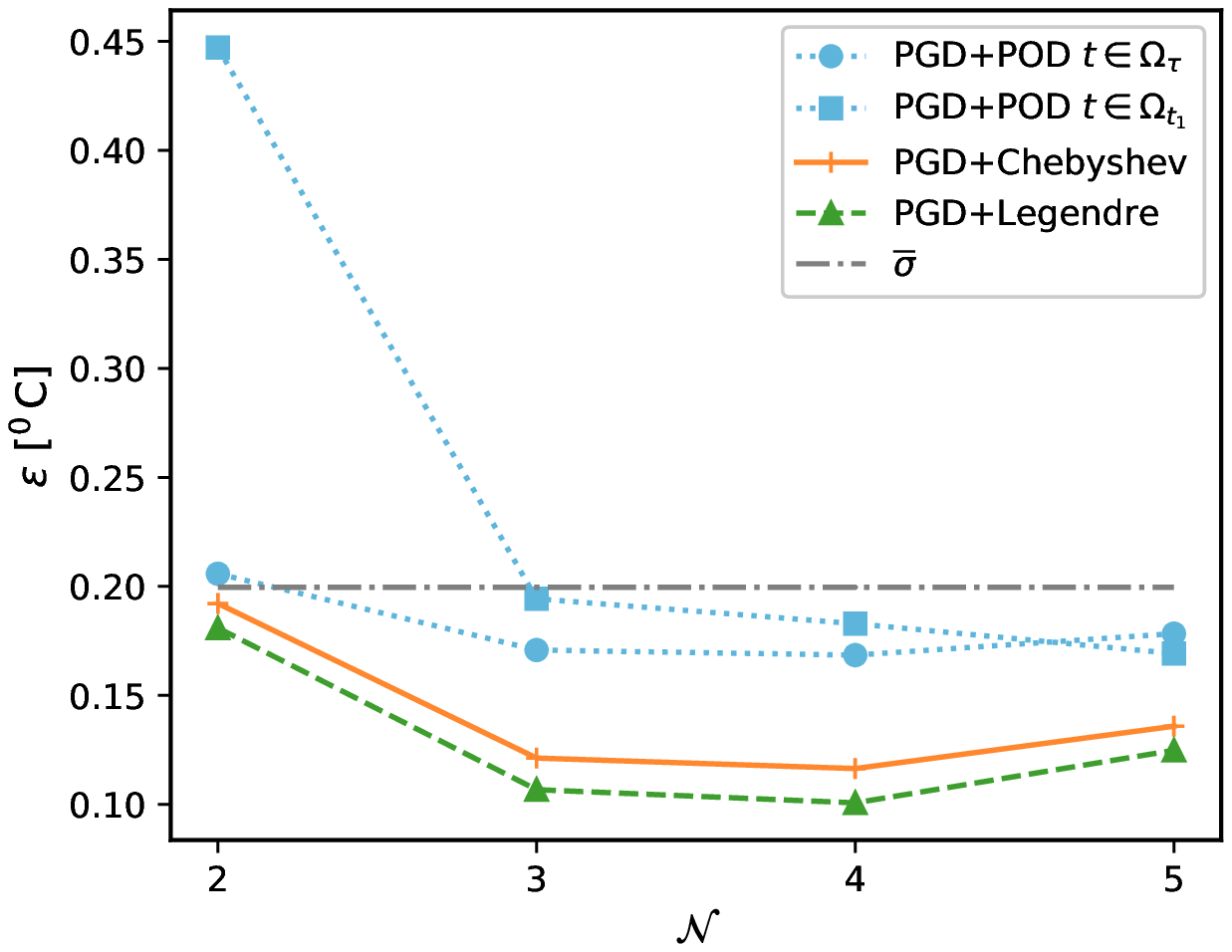}}
    \end{minipage}
        \hfill%
    \begin{minipage}[c]{.46\textwidth}
  \subfigure[For $\Delta \overline{\zeta} = 10^{-5}$, evolution of the $\varepsilon$ error at $8 \mathsf{cm}$ from the inner boundary condition]{\label{fig:Err_N_learning_period_Th04}\includegraphics[width=1\textwidth]{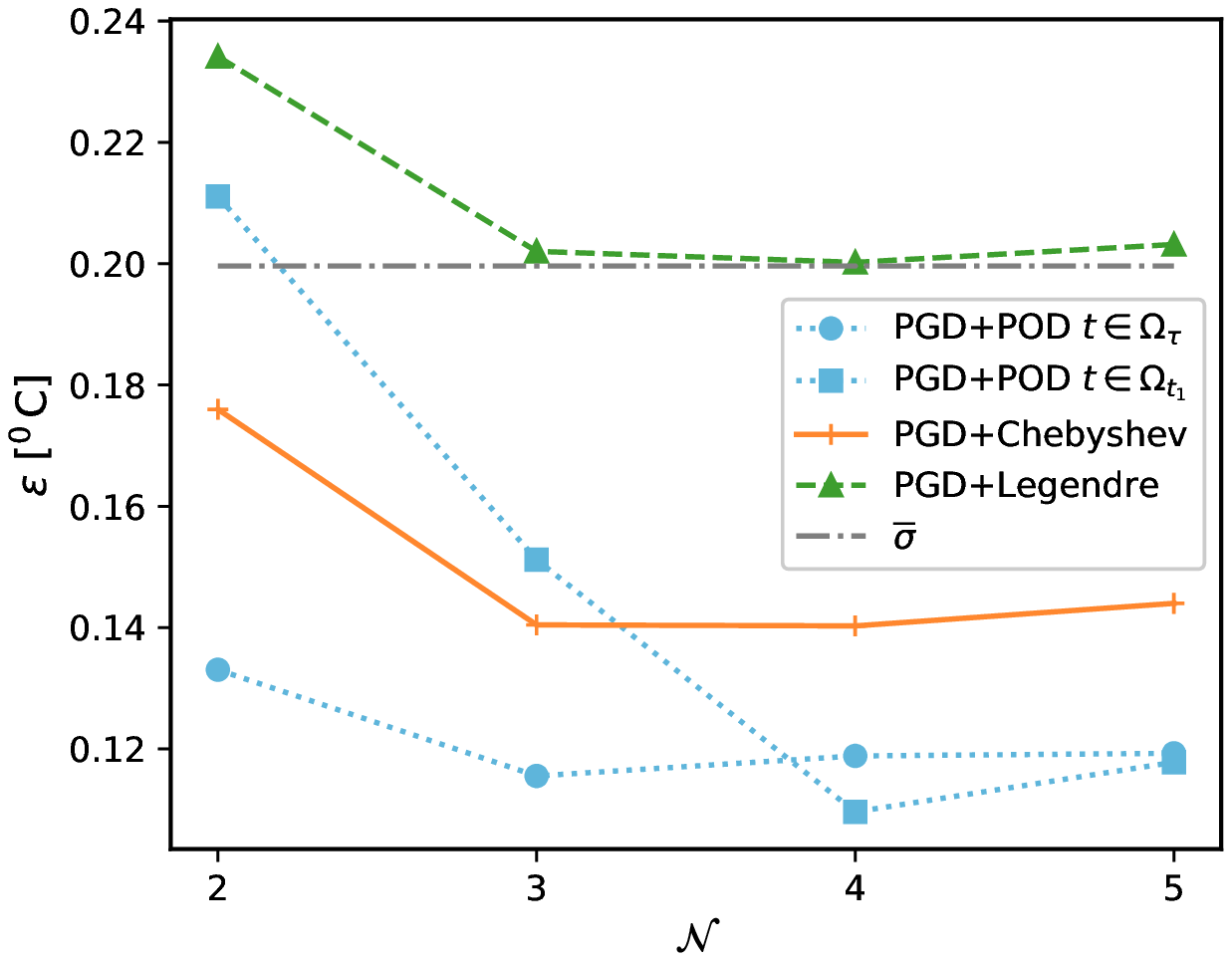}}
    \end{minipage}
  \caption{PGD parametric model $\varepsilon$ error to the measurement data ($\overline{\sigma}$ correspond to the mean experimental uncertainty)}
  \label{fig:error_PGD_TH03_Th04}
\end{figure}

Finally, the ability of the PGD parametric model to reproduce the dynamics on a realistic example is here studied. The results of the four models for $\mathcal{N}=3$ and $\Delta \overline{\zeta}= 10^{-5}$ are compared to the measurements. 
\bigbreak
Figure \ref{fig:Comp_expe_model} presents the time evolution at the position of sensors T02 and T03, respectively at $4 \; \mathsf{cm} $ and $8 \; \mathsf{cm} $ from the inner surface. All four models follow the dynamics of the measured curve. In figure \ref{fig:Th_03}, we can observe that the unfavorable POD basis matches the favorable POD basis for the first cycles, then the two curves depart from each other. It denotes the fact that the POD basis will be accurate as it encounters its training boundary conditions but will deviate as it encounters different boundary conditions.
\bigbreak
Figure \ref{fig:error_PGD_TH03_Th04} presents the error to the measurement data at both depth $4 \; \mathsf{cm} $ and $8 \; \mathsf{cm} $ from the inner boundary condition for the various numbers of modes $\mathcal{N}$. The same tendencies are observed as the ones described for figure \ref{fig:inf_learning period_PGD}. The error decreases and stabilizes after a few modes for each model. Depending on the reference data, $4 \; \mathsf{cm} $ and $8 \; \mathsf{cm}$, the method ranking is not the same. Results for $4 \; \mathsf{cm} $ are similar to the one observed comparing the PGD solution to the reference solution (finite difference model). In the results for $8 \; \mathsf{cm} $, we can see that the training period of the POD basis has less influence. Indeed at this location, the signal amplitude is eased. It fluctuates less. It could be easier to parameterize this part of the temperature profile.  
\bigbreak

This last study confirms the ability of the PGD parametric model with the approximation basis to reproduce the dynamics of the signal. At $4 \; \mathsf{cm} $, for $\mathcal{N} > 2$, all four models reach the mean experimental uncertainty. At $8 \; \mathsf{cm} $, for $\mathcal{N} > 2$, the \textsc{Legendre} combined models is getting closer to $\overline{\sigma}$, while the other models errors are under the threshold of the mean experimental uncertainty.
\bigbreak
Finally, the accuracy of the models is of the same order of magnitude than the reference solution. Indeed, if we quantify the error between the reference solution and the measurements, we obtain an error of $0.18\, \mathsf{\degC}$ at $4 \; \mathsf{cm}$ from the left boundary condition and $0.13\, \mathsf{\degC}$ at $8 \; \mathsf{cm}$ from the inner boundary condition. Those values are close to the one presented in Figure \ref{fig:error_PGD_TH03_Th04}. 

\section{Conclusions}
The POD, the \textsc{Chebyshev} and \textsc{Legendre} polynomial approximation basis have been compared first on a theoretical example. This case study was an opportunity to quantify the influence of three main parameters:

\begin{enumerate}
    \item the number of modes $\mathcal{N}$ in the approximation basis,
    \item the discretization coefficient,
    \item the number of modes $\mathcal{M}$ in the PGD basis.
\end{enumerate}

\bigbreak 
The different basis were then compared on a practical example based on measurements. This second case study intended to highlight the influence of the learning process on the accuracy of the POD basis. It also enables the comparison of the three combined PGD parametric models with measurements. 
\bigbreak 
The approximation basis have been first applied to the approximation of the source term. This first step has shown that the discretization should be selected in accordance with the number of modes $\mathcal{N}$. Indeed, increasing the number of modes with an insufficient discretization will not increase the accuracy of the approximation.
\bigbreak 
The different approximation basis were then integrated into the PGD parametric model. The first study on the influence of the discretization of the approximation coefficient revealed that the accuracy and the computation time are proportional to the discretization. The finer the mesh, the closer the discrete representation to the continuous function. However, as we increase the discretization, we increase the online calculation time. 
\bigbreak 
The study on the influence of the number of modes $\mathcal{N}$ has shown that the error decreases as we increase the number of modes in the approximation basis. This is not the case when the final accuracy is reached with a few modes as it was the case for the POD basis in the theoretical part. Finally, as the number of modes is increased, the computational time increases.
\bigbreak 
A relation has also been highlighted between the number of approximation modes $\mathcal{N}$ and the convergence rate of the fixed-point algorithm. As the number of modes increases, the number of parameters in the PGD model increases, decreasing the convergence rate of the algorithm. More modes $\mathcal{M}$ are then necessary for the PGD basis to achieve the same accuracy. 
\bigbreak 
The efficiency of the PGD parametric model depends on the three basis on the three previous parameters studied. A compromise should be found between the number of modes $\mathcal{N}$ and $\mathcal{M}$, the discretization and the computation time needed to compute and use the PGD combined model. 
\bigbreak 
The POD approximation basis has the main drawback to require a learning process. The benefit from a PGD parametric model as an \textit{a priori} method is then canceled out by the use of an \textit{a posteriori} method. The combined POD and PGD parametric model becomes then an \textit{a posteriori} model. Its performance depends on the training data-set used. 
\bigbreak 
The influence of this last parameter has been studied in the practical study case. Depending on the data-set used to train the POD basis, it could be the most or the less accurate method to parameterize the source term.
\bigbreak 
Finally, a compromise should be found to minimize the training period and the computational cost needed to build the basis while keeping an accurate approximation basis. For that purpose, a methodology to select an efficient training period should be developed. This is a point of current work. Some leads have been explored on how to improve the necessary training period by Berger \emph{et al.} (2018, \cite{berger2018intelligent}). A methodology has been proposed in \cite{azam2019learning} to select a short and representative training period for a building wall. 
\bigbreak 
As a final conclusion, we should keep in memory that the POD basis provides an optimum basis if the learning process is complete (the full data-set is used to build the basis). An efficient training data-set is then needed. However, when those data are not available, polynomial basis are a good alternative. They have the main benefit to provide an \textit{a priori} combined PGD parametric model.
\bigbreak 
However for both methods, the POD or the polynomial approximation, this work should be continued. For the POD method, the learning process remains the main barrier. For polynomial approximation, the parameterization of multi-material wall brings to light new questions. With a multi-layer wall, the source term may not be a smooth function. The efficiency of the polynomial basis to parameterize the temperature profile should then be tested. 
\bigbreak
Finally, the PGD model combined with each basis as shown its abilities to represent a realistic case study. Those models are ready to be aggregated with other sub-models through a co-simulation process to replace a large original model.

 \appendix

\renewcommand*{\thesection}{\Alph{section}}

\section{Details on the model error due to the inside radiative heat flux} 
\label{app:details_on_the_model_error}

As mentioned in Section~\ref{sec_2_1:Physical_problem}, the net radiative heat flux have been neglected on the inside part of the wall.  This heat flux is composed of the short and long-wave radiative heat flux. The short-wave radiative heat flux transmitted through the building windows is generally taken into account and distributed to the building interior surfaces (by solar tracking or with a weighted method) \cite{lauzet2019building}. For the long-wave radiative heat flux, it calculation requires the introduction of non-linear terms, most building simulation tools proposed then simplifications. This heat flux is either neglected, either linearised, and integrated into the convective heat transfer coefficient.

\subsection{Model error for the hypothesis neglecting the inside radiation effects}

To evaluate the impact of neglecting the inside net radiative heat flux, it is possible to propose a model error for this hypothesis. To obtain this model, the solution of the heat transfer equation considering inside radiation effects is denoted by $\widetilde{u}$. Then, the boundary condition on the inside part of the wall is: 
\begin{align*}
k \,  \frac{\partial \widetilde{u}}{\partial x} = - h_{\,\mathrm{in}} \, \bigl(\, \widetilde{u} - u_{\,\mathrm{in}} \,\bigr) + q_{\,\mathrm{in}} \,, \qquad x = L \,,
\end{align*}
where $q_{\,\mathrm{in}}$ is the incident radiation flux arising from the boundary surfaces facing the studied wall. The error between the solutions is defined by:
\begin{align}
\label{eq:def_error_model}
e \, \mathop{\stackrel{\,\mathrm{def}}{:=}\,} \, u - \widetilde{u} \,.
\end{align}
Recalling that $u$ is the solution of equation \ref{eq:HTE}, which neglect the inside net radiative heat flux. Since the problem is linear, the model error verifies the following governing equation:
\begin{align}
\label{eq:governing_eq_model_error}
c \, \frac{\partial e}{\partial t} = \frac{\partial}{\partial x} \,\biggl(\, k \, \frac{\partial e}{\partial x} \,\biggr) \,,
\end{align}
with the following boundary conditions: 
\begin{subequations}
\begin{align}
\label{eq:BC_model_error}
- k \,  \frac{\partial e}{\partial x}  & = - h_{\,\mathrm{out}} \, e \,, \qquad x = 0 \,,  \\[4pt]
k \,  \frac{\partial e}{\partial x}  & = - h_{\,\mathrm{in}} \, e - q_{\,\mathrm{in}} \,, \qquad x = L \, ,
\end{align}
\end{subequations}
and the initial condition:
\begin{align}
\label{eq:IC_model_error}
e = 0 \,, \quad t = 0 \,.
\end{align}
The model error equations ~\eqref{eq:governing_eq_model_error}--\eqref{eq:IC_model_error} can be computed using any of the numerical method presented in Section~\ref{sec_2_4:Approximation_basis} and Section~\ref{sec:PGD}. This is facilitated by working with dimensionless equations enabling to reuse the same numerical model for different problems.

\subsection{Results for the theoretical case study}

The use of the model error is illustrated for the case study defined in Section~\ref{sec:theoretical_case_study}.  The inside radiative heat flux is defined through long-wave radiation exchanges with surrounding surfaces:
\begin{align*}
q_{\,\mathrm{in}} = f_{\,\mathrm{w}} \, \epsilon_w \, \sigma \, 4 \, \bigl(\, T^{\,4} - u_{\,\mathrm{w}}^{\,4} \,\bigr)
+ f_{\,\mathrm{g}} \, \epsilon_g \, \sigma \, \bigl(\, T^{\,4} - u_{\,\mathrm{g}}^{\,4} \,\bigr)\,,
\end{align*} 
where $\sigma$ is the \textsc{Boltzmann} constant and $\epsilon_{w/g}$ the emissivity of the material. $u_{\,\mathrm{w}}$ and $u_{\,\mathrm{g}}$ are the surrounding walls and ground surface temperatures, respectively. The corresponding shape factor are $ f_{\,\mathrm{w}}$ and $ f_{\,\mathrm{g}}\,$. The first part of the formula corresponds to the radiative balance with the three walls and the ceiling, while the second part corresponds to the balance with the floor. 

\begin{figure}[ht!]
\begin{center}
\subfigure[\label{fig:q_ft}]{\includegraphics[width=.45\textwidth]{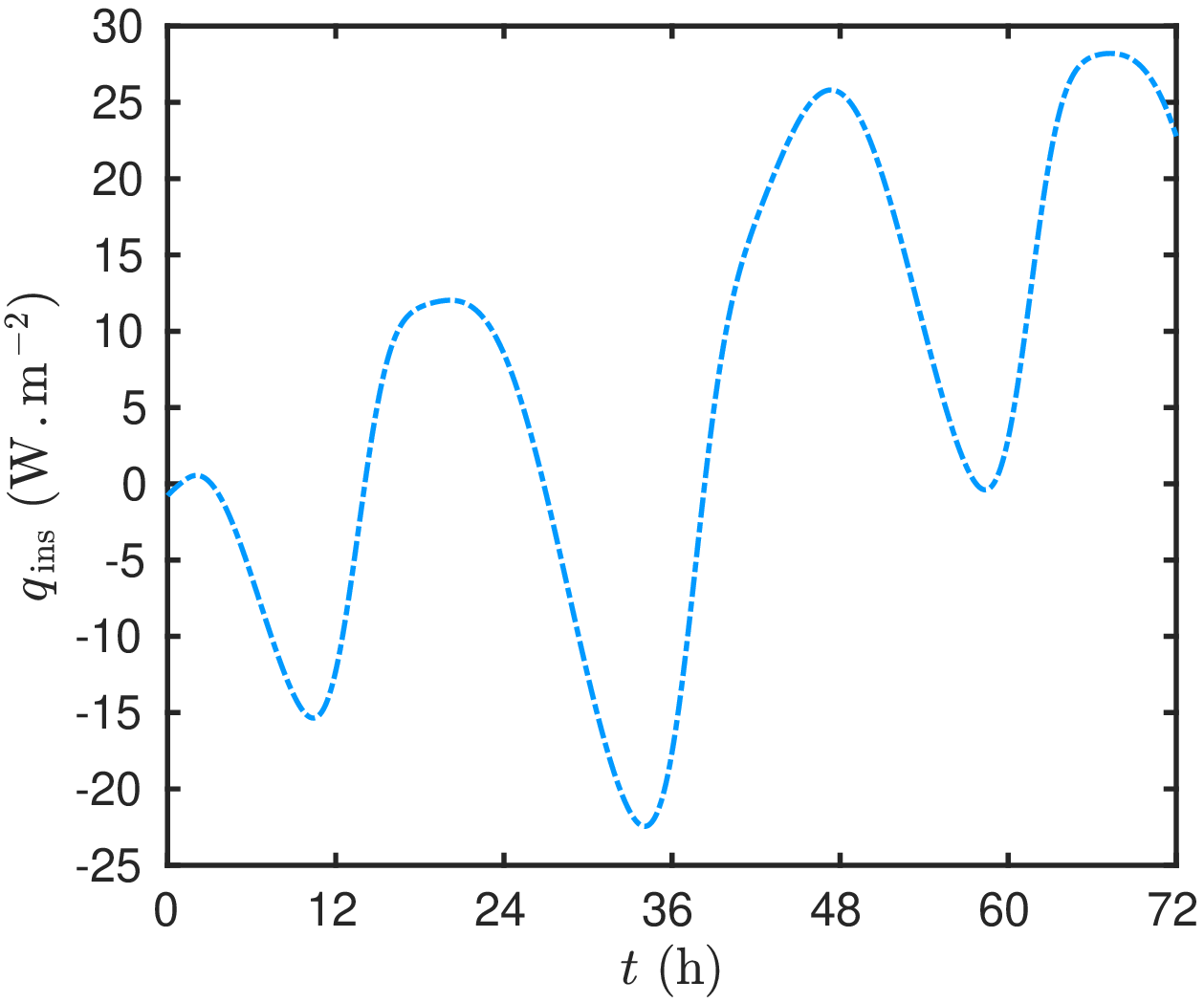}} \hspace{0.2cm}
\subfigure[\label{fig:e_ft}]{\includegraphics[width=.45\textwidth]{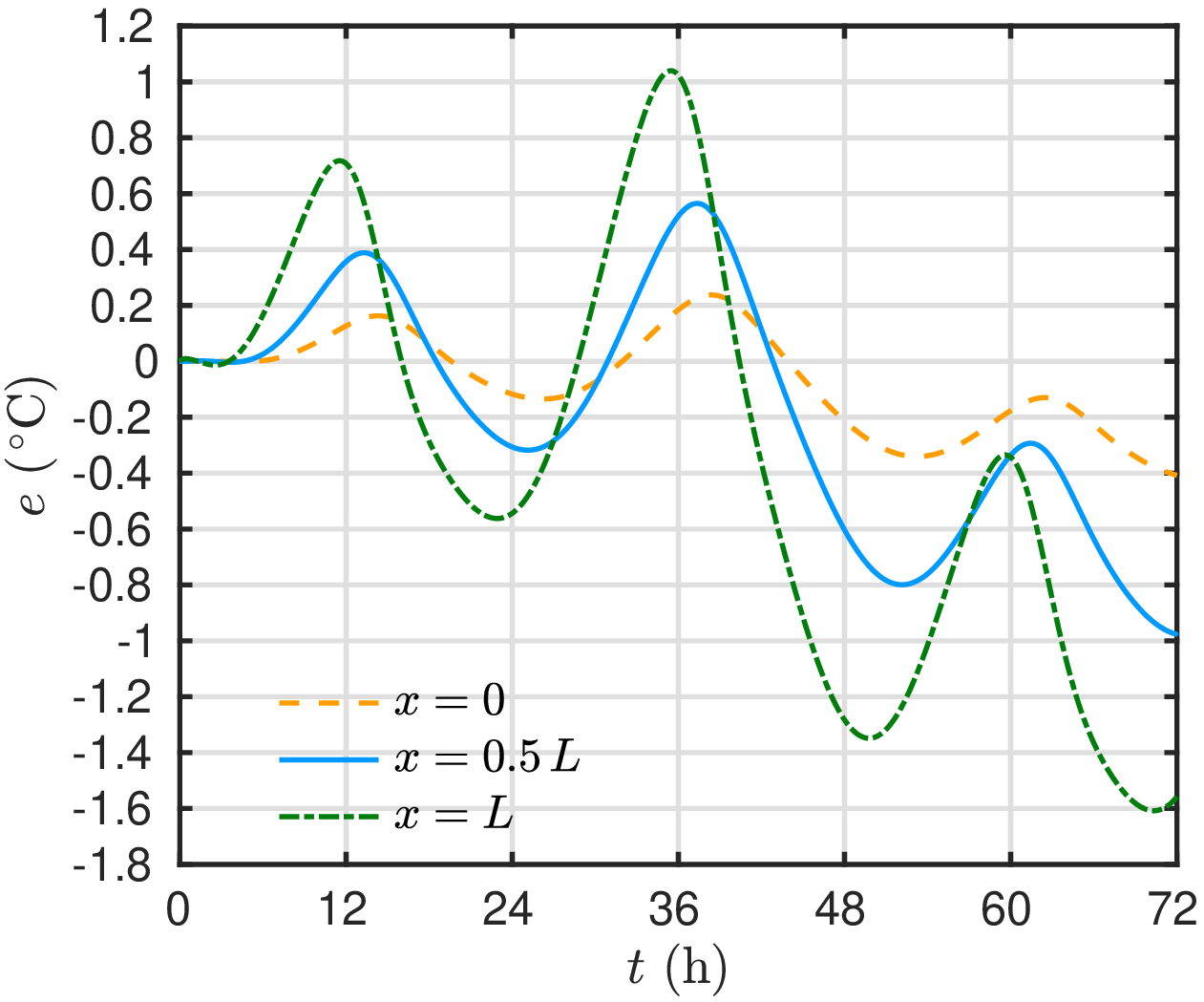}} 
\caption{Time evolution of the inside boundary flux due to long-wave radiation \emph{(a)} and of the model error \emph{(b)}}
\end{center}
\end{figure}

\begin{figure}[ht!]
\begin{center}
\subfigure[\label{fig:uxL_ft}]{\includegraphics[width=.45\textwidth]{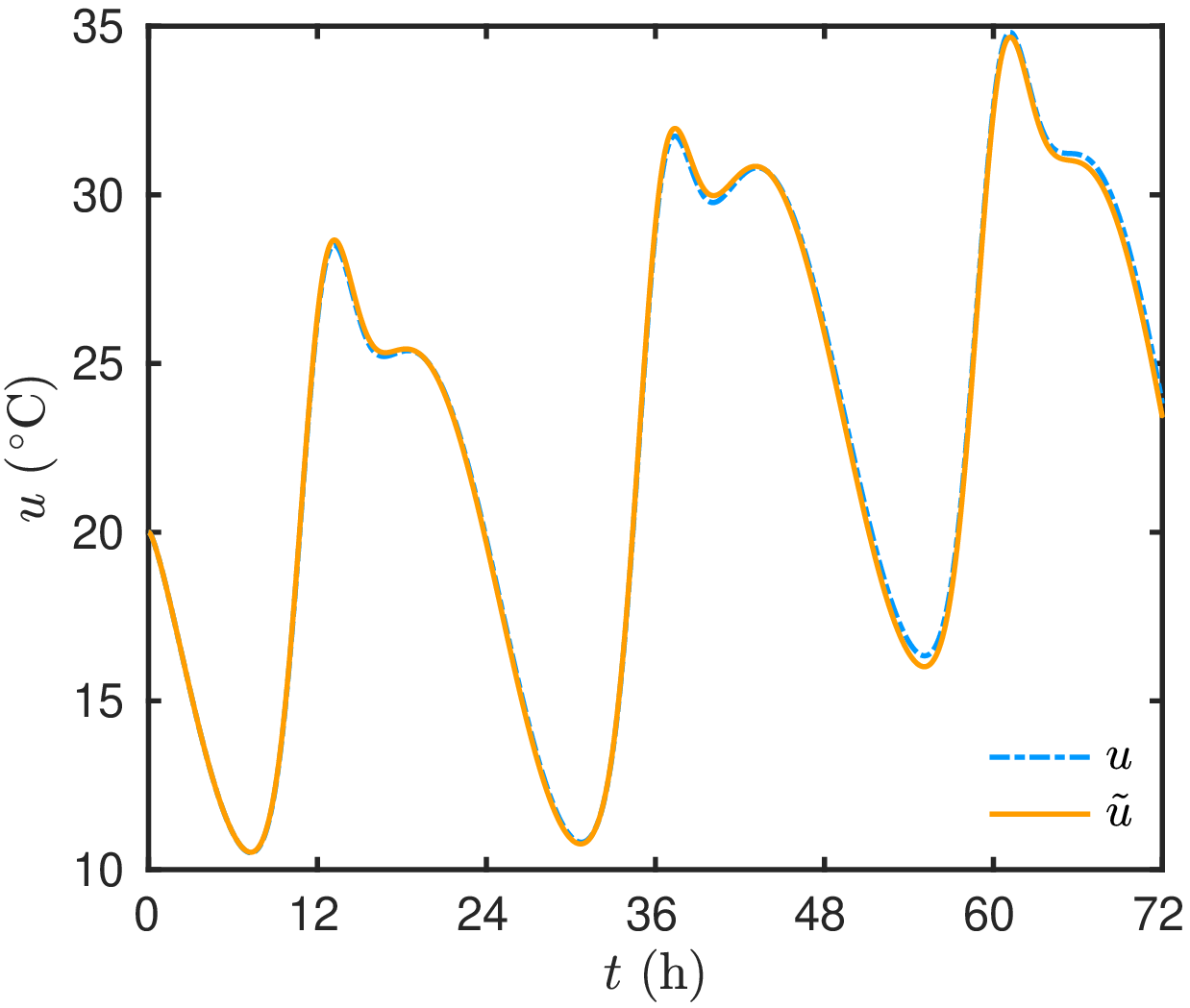}} \hspace{0.2cm}
\subfigure[\label{fig:uxR_ft}]{\includegraphics[width=.45\textwidth]{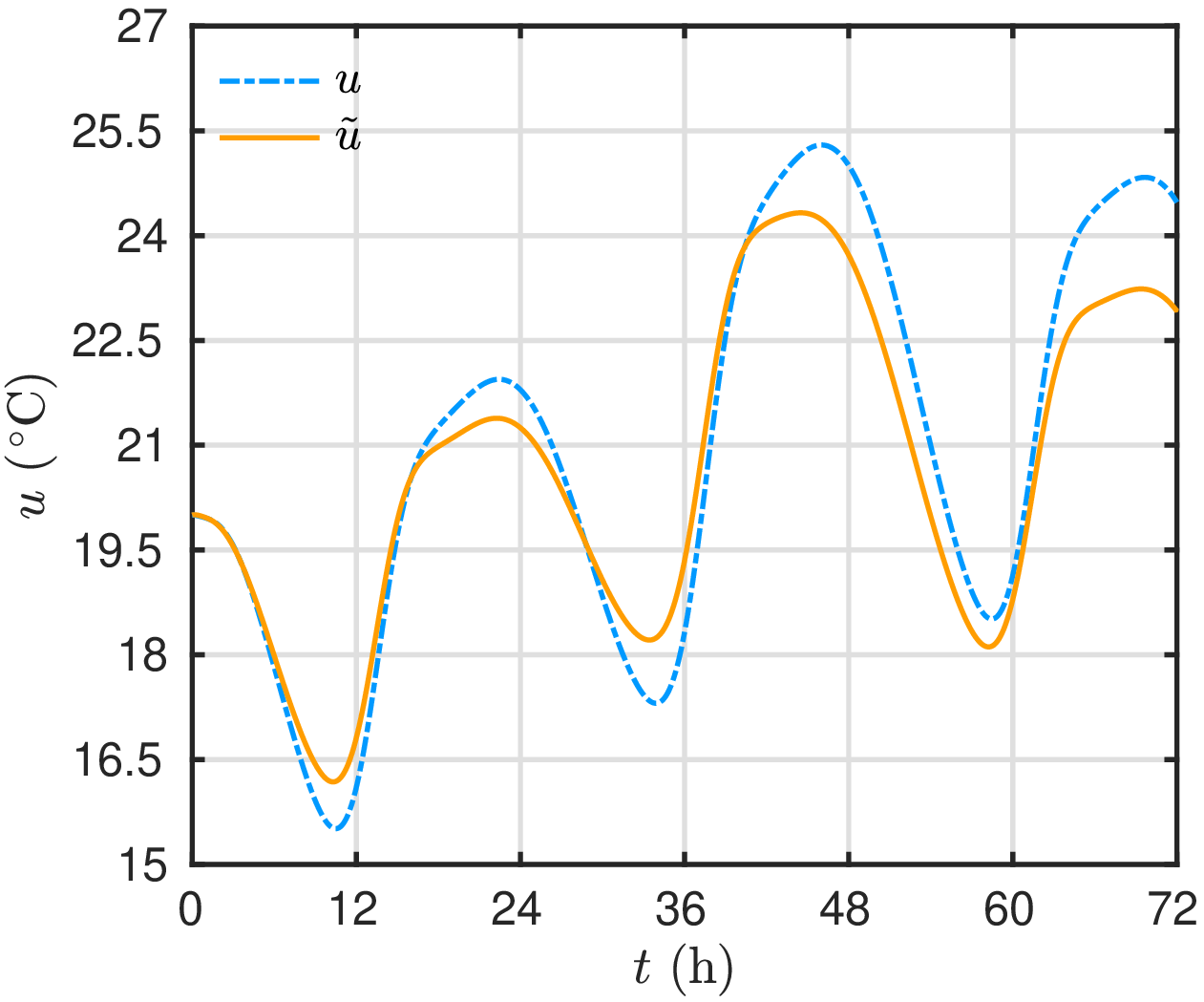}} 
\caption{Time evolution of the temperature with or without neglecting the inside radiation flux on the outside \emph{(a)} and inside \emph{(b)} surfaces.}
\end{center}
\end{figure}

For the numerical applications, the following values are considered: 
\begin{align*}
f_{\,\mathrm{w}} = f_{\,\mathrm{g}} = 0.2 \,, \,
& \epsilon_w = \epsilon_g = 0.9 \,, \,
& \sigma = 5.67 \cdot 10^{\,-8} \mathsf{W\,.\,m^{\,-2}\,.\,K^{\,-4}} \,, \,
& u_{\,\mathrm{w}} = u_{\,\mathrm{in}} \,, \,
& u_{\,\mathrm{g}} = 23 \ \mathsf{^{\,\circ}C} \,.
\end{align*}
To obtain the previous numerical values, the followings hypothesis have been made: - the room studied has no windows, - the room is perfectly cubic (all the shape factors are equal to 0.2), -  the surface temperatures of the walls and ceiling are equal to the air temperature (an equilibrium has been reached with neighboring rooms), - the floor surface temperature equal to $23\, \mathsf{\degC}$ (underfloor heating). 
\bigbreak
The flux $q_{\,\mathrm{in}}$ is computed using \emph{a posteriori} results of the wall. The time variation of the flux is shown in Figure~\ref{fig:q_ft}. It can be remarked that the radiation flux scales between $-25$ and $30 \ \mathsf{W\,.\,m^{\,-2}}\,$. It has a very low magnitude compared to the outside flux, illustrated in Figure~\ref{fig:qe}. Using the time variation of $q_{\,\mathrm{in}}\,$, the model error is computed based on a finite-difference model. The time variation of the model error is given in Figure~\ref{fig:e_ft}. The error reaches a maximum of $1.0 \ \mathsf{^{\,\circ}C}$ located, as expected, on the inside boundary ($x= L$). The impact of the hypothesis neglecting the inside flux can be evaluated on the temperature flux. For this, the solution $\widetilde{u}$ is reconstructed using Eq.~\eqref{eq:def_error_model}. The temperature variation are illustrated in Figures~\ref{fig:uxL_ft} and \ref{fig:uxR_ft}. On the outside surface, the two solutions are almost overlapped. Thus, the influence of the inside radiation is negligible on this part. Indeed, as remarked in Figure~\ref{fig:e_ft}, the model error scales with $0.2 \ \mathsf{^{\,\circ}C}\,$. On the inside surface, the discrepancy between the solution is  higher, around $0.5 \ \mathsf{^{\,\circ}C}\,$. 
\bigbreak
As a synthesis, a model error is proposed to evaluate the influence of the hypothesis neglecting the inside net radiative heat flux. It can be computed using any of the numerical models proposed in the manuscript, due to the benefits of working with dimensionless equations. In terms of physical results, the inside radiation effects induce discrepancies on the inside surface of the wall. However, the overall dynamics of heat transfer is not altered. Note that the numerical investigations carried in Sections~\ref{sec:theoretical_case_study} 
 can be straightforwardly extended to a model considering inside radiation flux.

\section{Details on the approximation basis construction} 
\label{app:details_on_the_app_basis_cons}

\subsection{\textsc{Chebyshev} polynomials}
\label{sec_app:details_on_Chebyshev}

 The \textsc{Chebyshev} polynomials are part of the family of orthogonal polynomials. The first kind \textsc{Chebyshev} polynomial denoted $T_{n}$ are the following ones: 

\begin{equation}
    T_{\,0} \bigl(\,x\,\bigr) \ = \ 1,  \; T_{\,1} \bigl(\,x\,\bigr) \ = \ x, \; T_{\,2} \ = \ \bigl(\,2\, x^2\ - \ 1\,\bigr), \; T_{\,3} \ = \ \bigl(\,4\, x^3\ - \ 3\, x\,\bigr).
\end{equation}

They are constructed according to the following relation of recurrence  \cite{peyret2013spectral,trefethen2013approximation}: 

\begin{equation}
    T_{\, j \ + \ 1} \ =  \ 2 \, x \ T_{\, j} \ - \ T_{\, j \ - \ 1} \; \text{for} \; j\  > \ 1 \ \text{with}  \
    T_{0}=1 ,\ \text{and},\
    T_{1}=x
\end{equation}

The \textsc{Chebyshev} approximation basis is made of the \textsc{Chebyshev} polynomials.

\begin{equation}
    \Psi_{\, j} \ \equiv \ T_{\, j} 
\label{eq:Chebyshev_basis}
\end{equation}

The \textsc{Chebyshev} polynomials are calculated at the \textsc{Chebyshev} points defined by the equation  \eqref{eq:Chebyshev_points}, where $n$ is a positive integer.  In the literature several names can be found to describe this set of points as \textsc{Chebyshev–Lobatto} points, \textsc{Chebyshev}  extreme points, or  \textsc{Chebyshev}  points of the second kind. All those expressions refer to the same set of points according to Trefethen (2013, \cite{trefethen2013approximation}). 

\begin{equation}
x_{\, j} \ = \ \cos \left(\, \frac{j \, \pi}{n}  \right)\, , \; \;  0 \ < \ j \ < \ n,
\label{eq:Chebyshev_points}
\end{equation}

Special attention must be given to the spatial domain of the problem. The \textsc{Chebyshev} points define a non-uniform mesh for a space interval $[-1,1]$. Thus, a change of variable must be performed to transform the dimensionless spatial domain $[0,1]$ to $x \in [-1,1]$. 
\subsection{\textsc{Legendre} polynomials}
\label{sec_app:details_on_Legendre}

The Legendre polynomials are also part of the family of orthogonal polynomials. The first \textsc{Legendre} polynomials are the following ones:
\begin{equation}
    P_{\,0} \bigl(\,x\,\bigr) \ = \ 1,  \; P_{\,1} \bigl(\,x\,\bigr) \ = \ x, \; P_{\,2} \ = \ \left(\, \frac{3}{2} \, x^2\ - \ \frac{1}{2}\,\right).
\end{equation}
The next polynomials are constructed according to the following relation of recurrence \cite{trefethen2013approximation}: 
\begin{equation}
 \left( \,  j \ + \ 1 \,\right) \,  P_{\, j \ + \ 1} \ =  \ ( \, 2\,  j \ + \ 1 \,)\, x \, P_{\, j} \ - \ j \, P_{\, j \ - \ 1} \; \text{for} \; j \  \geq \ 1, \ \text{and} \ P_{\,0} \bigl(\,x\,\bigr) \ = \ 1 \ , \  P_{\,1} \bigl(\,x\,\bigr) \ = \ x
\end{equation}
The \textsc{Legendre} approximation basis is made of the \textsc{Legendre} polynomials calculated at the \textsc{Legendre} points.
\begin{equation}
    \Psi_{\, j}  \ \equiv \ P_{\, j} 
\label{eq:Legendre_basis}
\end{equation}
As for \textsc{Chebyshev}, special attention must be given to the spatial domain. The spatial mesh will not be uniform and a change of variable must be performed to transform the dimensionless spatial domain from $[0,1]$ to  $x \in  [-1,1]$.

\subsection{POD reduced basis}
\label{sec_app:details_on_POD_method}
The POD method extracts the relevant information from a set of \textit{snapshots} by means of its projection onto a smaller subspace. As a result, from a data-set, the POD builds a deterministic representation, from the basis $\Phi$. The ultimate goal is to retain a detailed representation of the data-set with a minimum or optimal number of modes in $\Phi$. For these properties, the POD method could be used to parameterize the temperature profile (source term in our problem).
\begin{equation}
    \Psi_{\, j}  \ \equiv \ \Phi_{\, j} 
\label{eq:POD_basis}
\end{equation}
To build the POD basis, a learning process is needed. It has an impact on the accuracy of the reduced-order basis. For this reason, the data-set used must be representative of the problem (boundary values, initial conditions, materials used). More details on the POD methods can be found in  \cite{liang2002proper,cueto2014model}.
\bigbreak
Contrary to the two previous basis, no special attention needs to be paid to the definition of the spatial domain. To standardize the spatial domain used, the same change of variable is performed ($x \in  [-1,1]$) and the spatial mesh is set uniform.

\bibliographystyle{unsrt}
\bibliography{sample}

\end{document}